\documentclass[12pt]{article}
\usepackage{epsfig,cite,amsmath}

\input paperdef

\def\_{\rule{.3em}{.15ex}}

\def\slash#1{\setbox0=\hbox{$#1$}#1\hskip-\wd0\dimen0=5pt\advance
       \dimen0 by-\ht0\advance\dimen0 by\dp0\lower0.5\dimen0\hbox
         to\wd0{\hss\sl/\/\hss}}

\allowdisplaybreaks

\begin{document}
\thispagestyle{empty}

\def\thefootnote{\fnsymbol{footnote}}

\begin{flushright}
DCPT/02/110\\
IPPP/02/55\\
LMU 07/02\\
hep-ph/0209305 \\
\end{flushright}

\vspace{1cm}

\begin{center}

{\large\sc {\bf Leading Electroweak Two-Loop Corrections}}

\vspace{0.4cm}

{\large\sc {\bf to Precision Observables in the MSSM}}
 
\vspace{1cm}

{\sc 
S.~Heinemeyer$^{1}$%
\footnote{email: Sven.Heinemeyer@physik.uni-muenchen.de}%
~and G.~Weiglein$^{2}$%
\footnote{email: Georg.Weiglein@durham.ac.uk}
}

\vspace*{1cm}

{\sl
$^1$Institut f\"ur theoretische Elementarteilchenphysik,
LMU M\"unchen, Theresienstr.\ 37, D-80333 M\"unchen, Germany

\vspace*{0.4cm}

$^2$Institute for Particle Physics Phenomenology,\\ 
University of Durham, Durham DH1 3LE, U.K.
}

\end{center}

\vspace*{0.2cm}

\begin{abstract}
The leading electroweak MSSM \twol\ corrections to the
electroweak precision observables are calculated. They are 
obtained by evaluating the \twol\ 
\order{\al_t^2}, \order{\al_t \al_b}, \order{\al_b^2} contributions to
the quantity~$\De\rho$ in the limit of heavy scalar quarks, i.e.\ we
consider the contributions of a Two-Higgs-Doublet model with MSSM
restrictions. The full analytic result
for arbitrary values of the lightest $\cp$-even Higgs boson mass is
presented.
The numerical effects of the leading electroweak MSSM \twol\ corrections
on the precision observables $\MW$ and $\sweff$
are analyzed. The electroweak \twol\ contribution 
to $\MW$ amounts up to $-12 \mev$ and up
to $+6 \times 10^{-5}$ for $\sweff$.
The corrections from the bottom quark loops can become important for
large values of $\tb$. They enter with a different sign than the
\order{\al_t^2} corrections. 
We furthermore investigate the current sensitivity of the electroweak
precision observables to the top Yukawa coupling in the SM and the
MSSM. The prospects for indirectly determining this coupling at the next
generation of colliders are discussed. 
\end{abstract}

\def\thefootnote{\arabic{footnote}}
\setcounter{page}{0}
\setcounter{footnote}{0}

\newpage


\section{Introduction}

Theories based on Supersymmetry (SUSY) \cite{susy} are widely considered as the
theoretically most appealing extension of the Standard Model (SM). They are
consistent with the approximate unification of the gauge coupling
constants at the GUT scale and provide a way to cancel the quadratic
divergences in the Higgs sector hence stabilizing the huge hierarchy between
the GUT and the Fermi scales. Furthermore, in SUSY theories the breaking
of the electroweak symmetry is naturally induced at the Fermi scale,
and the lightest supersymmetric particle can be neutral, weakly interacting
and absolutely stable, providing therefore a natural solution for the
Dark Matter problem.

Supersymmetry predicts the existence of scalar partners $\tilde{f}_L, 
\tilde{f}_R$ to each SM chiral fermion, and spin--1/2 partners to the 
gauge bosons and to the scalar Higgs bosons. So far, the direct search for
SUSY particles has not been successful. 
One can only set lower bounds of ${\cal O}(100)$~GeV on 
their masses~\cite{pdg}. 
Furthermore, contrary to the SM two Higgs doublets are
required resulting in five physical Higgs bosons~\cite{hhg}. The
direct search resulted in lower limits of about $90 \gev$ for the
neutral Higgs bosons and about $80 \gev$ for the charged
ones~\cite{lephiggs}. 

An alternative way to probe SUSY is via the virtual effects of the 
additional particles to precision observables. 
This requires a very high precision 
of the experimental results as well as of the theoretical predictions.
The most prominent role in this respect plays the
$\rho$-parameter~\cite{rho}. The radiative corrections from vector boson
self-energies to the quantity $\De\rho$ constitute the leading,
process independent corrections 
to many electroweak precision observables, such as the prediction for
$\De r$, i.e.\ the $\MW-\MZ$~interdependence, 
and the effective leptonic weak mixing angle, $\sweff$. 

The radiative corrections to the electroweak precision observables
within the Minimal Supersymmetric Standard Model (MSSM) 
stemming from scalar fermions, charginos, neutralinos and Higgs bosons 
have been discussed at the \onel\ level in~\citeres{dr1lA,dr1lB},
providing the full \onel\ corrections.
More recently also the leading \twol\ corrections in $\oaas$ to the
quark and scalar quark loops for $\De\rho$ 
have been obtained~\cite{dr2lA} as well as the gluonic \twol\ corrections
to the $\MW-\MZ$~interdependence~\cite{dr2lB}. Contrary to the SM
case, these \twol\ corrections turned out to increase the \onel\
contributions, leading to an enhancement of up to 35\%~\cite{dr2lA}.

In this paper we present the leading \twol\ corrections to 
$\De\rho$ at \order{\al_t^2}, \order{\al_t \al_b}, \order{\al_b^2},
i.e.\ the leading two-loop contributions involving the top and bottom
Yukawa couplings. 
These contributions are of particular interest, since they involve
corrections proportional to $\mt^4$ and bottom loop corrections
enhanced by $\tb$, the ratio of the two vacuum expectation values. For
a large SUSY scale, $\msusy \gg \MZ$, the  
contributions from loops of SUSY particles decouple from physical
observables~\cite{decoupling1l,dr2lA}. 
Therefore, focusing on the case of large $\msusy$, we derive the
leading electroweak  \twol\ corrections in the limit where besides the SM 
particles only the two Higgs doublets of the MSSM are active. 

As a first step, in \citere{drMSSMgf2} we have calculated the
\order{\al_t^2} corrections in the limit where the lightest $\cp$-even
Higgs boson mass vanishes, i.e.\ $\mh \to 0$. The numerical
effect of these corrections turned out to be relatively
small. However, for the corresponding SM result it was found that the
$\MHSM = 0$ limit is only a poor approximation of the result with
arbitrary $\MHSM$~\cite{drSMgf2mt4}. Since a similar behavior can be
expected for the MSSM, we perform the calculation of the leading
electroweak \twol\ corrections, 
\order{\al_t^2}, \order{\al_t \al_b}, and \order{\al_b^2}, 
for arbitrary values of $\mh$. 
The result obtained in the MSSM is compared with the corresponding SM
correction of \order{\al_t^2}~\cite{drSMgf2mt4}. The resulting
shift in $\MW$ and $\sweff$ is analyzed numerically.

Since the top Yukawa coupling enters the predictions for the
electroweak precision observables at lowest order in the perturbative
expansion at
\order{\al_t^2}, these contributions allow to study the sensitivity of
the precision observables on this coupling. 
Using a simple approach in which we treat the top Yukawa coupling in
the SM and the MSSM as a free parameter, we study the current
sensitivities of the electroweak precision observables as well as the
prospective accuracies at the next generation of colliders. 

The rest of the paper is organized as follows: in Sect.~2 we review
the SM and MSSM corrections to the quantity $\De\rho$ and present the details
of the calculation of the \order{\al_t^2}, \order{\al_t \al_b},
\order{\al_b^2} corrections.
Explicit formulas for the results of \order{\al_t^2}, 
\order{\al_t \al_b}, and \order{\al_b^2} can be found in Sect.~3 and
the appendix. The numerical analysis is performed in Sect.~4. 
In Sect.~5 we analyze the sensitivity of the electroweak precision
observables to the top Yukawa coupling. 
We conclude with Sect.~6.


\section{Calculation of the \boldmath{\order{\al_t^2}, \order{\al_t \al_b}, 
          and \order{\al_b^2}}\\ corrections}
\label{sec:calc}

\subsection{One-loop results}
\label{subsec:oneloop}

The quantity $\De\rho$, 
\BE
\De\rho = \frac{\Si_Z(0)}{\MZ^2} - \frac{\Si_W(0)}{\MW^2} ,
\label{delrho}
\EE
 parameterizes the leading universal corrections to the electroweak
 precision observables induced by
the mass splitting between fields in an isospin doublet~\cite{rho}.
$\Si_{Z,W}(0)$ denote the transverse parts of the unrenormalized $Z$
 and $W$ boson self-energies at zero momentum transfer, respectively.
$\De\rho$ gives the dominant contribution to electroweak
precision observables, such as the $W$ boson mass, $\MW$, and the
effective leptonic mixing angle, $\sweff$. The induced shifts are in
 leading order given by (with $1-\sw^2 = \cw^2 = \MW^2/\MZ^2$)
\BE
\de\MW \approx \frac{\MW}{2}\frac{\cw^2}{\cw^2 - \sw^2} \De\rho, \quad
\de\sweff \approx - \frac{\cw^2 \sw^2}{\cw^2 - \sw^2} \De\rho .
\label{precobs}
\EE

In the SM the dominant contribution to $\De\rho$ at the \onel\ level
is given by the $t/b$ doublet due to their large mass splitting.
It reads
\BE
\De\rho^{\SM}_0 = \frac{3\,\gf}{8\,\wz\,\pi^2} \; F_0(\mt^2, \mb^2) ,
\label{delrhoSM1l}
\EE
with
\BE
F_0(x,y) = x + y - \frac{2\,x\,y}{x - y} \log \frac{x}{y} .
\EE
$F_0$ has the properties $F_0(m_a^2, m_b^2) = F_0(m_b^2, m_a^2)$, 
$F_0(m^2, m^2) = 0$, $F_0(m^2, 0) = m^2$. Therefore for $\mt \gg \mb$
\refeq{delrhoSM1l} reduces to the well known quadratic correction in $\mt$,
\BE
\De\rho^{\SM}_0 = \frac{3\,\gf}{8\,\wz\,\pi^2} \; \mt^2 .
\EE

Within the MSSM the dominant correction from SUSY particles at the
\onel\ level arises from 
the scalar top and bottom contribution to \refeq{delrho}. For 
$\mb \neq 0$ it is given by
\BEA
\De\rho_0^\SU = \frac{3\,\gf}{8\, \wz\, \pi^2} 
                  \Big[ -\sinQtt\cosQtt F_0(\mste^2, \mstz^2)
                       -\sinQtb\cosQtb F_0(\msbe^2, \msbz^2) \non\\
                       +\cosQtt\cosQtb F_0(\mste^2, \msbe^2)
                       +\cosQtt\sinQtb F_0(\mste^2, \msbz^2) \non\\
                       +\sinQtt\cosQtb F_0(\mstz^2, \msbe^2)
                       +\sinQtt\sinQtb F_0(\mstz^2, \msbz^2) ~\Big] .
\label{delrhoMSSM1l}
\EEA
Here $\msti, \msbi (i = 1,2)$ denote the stop and sbottom masses,
whereas $\tst, \tsb$ are the mixing angles in the stop and in the
sbottom sector.


\subsection{Results beyond the \onel\ level}
\label{subsec:beyondoneloop}

Within the SM the \onel\ \order{\al} result has been extended in
several ways. The dominant \twol\ corrections at
$\oaas$ are given by~\cite{drSMgfals}
\BE
\De\rho^{\SM, \al\als}_1 = - \De\rho^{\SM}_0 \frac{2}{3} \frac{\als}{\pi}
                           \KL 1 + \pi^2/3 \KR .
\EE
These corrections screen the \onel\ result by approximately 10\%. Also
the three-loop result at \order{\al\als^2} is known. Numerically it
reads~\cite{drSMgfals2}
\BE
\De\rho^{\SM, \al\als^2}_2 = -\frac{3\,\gf}{8\,\wz\,\pi^2} \, \mt^2 
                             \KL \frac{\als}{\pi} \KR^2 \cdot 14.594... ~.
\EE
Furthermore the leading electroweak \twol\ top quark contributions of
\order{\al_t^2} have been calculated. They enter the electroweak
precision observables together with the one-loop contribution
according to 
\BE
\rho = \frac{1}{1 - \De\rho}, \quad \De\rho = \De\rho_0 + \De\rho_1 .
\EE
First the result for $\De\rho_1$ in the limit
$\MHSM = 0$ had been evaluated~\cite{drSMgf2mh0},
\BEA
\De\rho^{\SM,\al_t^2}_{1|\MH = 0} &=&
 3\,\frac{\gf^2}{128 \pi^4} \, \mt^4\, \cdot \de^{\SM}_{1|\MH = 0} \non \\
\de^{\SM}_{1|\MH = 0} &=& 19 - 2 \pi^2 .
\EEA
Later the full \order{\al_t^2} result without restrictions in the
Higgs boson mass became available~\cite{drSMgf2mt4}, where 
$\de^{\SM}_{1|\MH = 0}$ extends to
\BE
\de^{\SM}_{1|\MH \neq 0} = 19 - 2 \pi^2 + {\rm fct}(\mt, \MH)~.
\label{drSMgf2}
\EE
Here ${\rm fct}(\mt, \MH)$ contains the extra terms arising from a
non-vanishing Higgs boson mass. 
Recently also first electroweak three-loop results in the limit of
$\MH = 0$ became available~\cite{drSMgf3mh0}. Numerically they read
\BEA
\De\rho^{\SM,\al_t^3}_{2|\MH = 0} &=& 
\KL \frac{\gf}{8\,\wz\,\pi^2} \mt^2 \KR^3 \cdot 249.74 ~, \\
\De\rho^{\SM,\al_t^2\als}_{2|\MH = 0} &=& 
\KL \frac{\gf}{8\,\wz\,\pi^2} \mt^2 \KR^2 
\KL \frac{\als}{\pi} \KR \cdot 2.9394 ~.
\EEA

\bigskip
In the MSSM up to now the \twol\ calculations have been restricted to
the leading $\oaas$ corrections to the scalar quark
loops~\cite{dr2lA}. They consist of the rather lengthy result
for gluino exchange, which decouples for $\mgl \to \infty$, and of the
compact correction for the gluon exchange contribution~\cite{dr2lA}:
\BEA
\De\rho^\SU_{1,{\rm gluon}} 
     = \frac{\gf}{4\, \wz\, \pi^2} \frac{\als}{\pi}
                 \Big[ -\sinQtt\cosQtt F_1(\mste^2, \mstz^2)
                       -\sinQtb\cosQtb F_1(\msbe^2, \msbz^2) \non\\
                       +\cosQtt\cosQtb F_1(\mste^2, \msbe^2)
                       +\cosQtt\sinQtb F_1(\mste^2, \msbz^2) \non\\
                       +\sinQtt\cosQtb F_1(\mstz^2, \msbe^2)
                       +\sinQtt\sinQtb F_1(\mstz^2, \msbz^2) ~\Big] ,
\label{delrhoMSSM2l}
\EEA
with
\BEA
F_1(x, y) &=& x + y - 2\frac{x y}{x - y} 
          \log\frac{x}{y} \KKL 2 + \frac{x}{y}\ln\frac{x}{y} \KKR \non \\
 &&         + \frac{(x + y)x^2}{(x - y)^2} \log^2\frac{x}{y}
            -2 (x - y) {\rm Li}_2 \KL 1 - \frac{x}{y} \KR, 
\EEA
where $F_1$ has the properties $F_1(m_a^2, m_b^2) = F_1(m_b^2, m_a^2)$, 
$F_1(m^2, m^2) = 0$, $F_1(m^2, 0) = m^2 (1 + \pi^2/3)$.

Contrary to the SM case where the strong \twol\ corrections screen the
\onel\ result, the $\oaas$ corrections in the MSSM increase the \onel\
contributions by up to 35\%, thus enhancing the sensitivity to scalar
quark effects. 
Another difference between the SM and the MSSM is the $\mt$
dependence of the leading contribution to $\De\rho$
within the SM. They are $\sim \mt^2$ for the \onel\ and $\sim \mt^4$
for the \twol\ correction leading to sizable shifts to the precision
observables. Concerning the corrections from loops of SUSY particles,
on the other hand, no large prefactor $\sim\mt^2$ at the \onel\ level
is present.
This behavior changes with the
leading electroweak \twol\ SUSY corrections, which are $\sim \mt^4$,
i.e.\ of \order{\al_t^2}. Therefore we concentrate on these and the
corresponding \order{\al_t \al_b}, \order{\al_b^2} corrections in this
paper. 
Since the SUSY loop contributions in the MSSM decouple if the general
soft SUSY-breaking scale goes to infinity, $\msusy \to
\infty$~\cite{decoupling1l,dr2lA}, the leading contributions for large
$\msusy$ arise from a Two-Higgs-Doublet model with MSSM restrictions.


\subsection{The Higgs sector of the MSSM}
\label{subsec:mssmhiggs}

Contrary to the SM, in the MSSM two Higgs doublets
are required~\cite{hhg}.
At the tree-level, the Higgs sector can be described with the help of two  
independent parameters (besides $g$ and $g'$): the ratio of the two
vacuum expectation values,  
$\tb = v_2/v_1$, and $M_A$, the mass of the $\cp$-odd $A$ boson.
The diagonalization of the bilinear part of the Higgs potential,
i.e.\ the Higgs mass matrices, is performed via orthogonal
transformations with the angle $\al$ for the $\cp$-even part and with
the angle $\be$ for the $\cp$-odd and the charged part.
The mixing angle $\al$ is determined through
\BE
\tan 2\al = \tan 2\be \; \frac{\MA^2 + M_Z^2}{\MA^2 - M_Z^2} ;
\qquad  -\frac{\pi}{2} < \al < 0~.
\label{alphaborn}
\EE
One gets the following Higgs spectrum:
\BEA
\mbox{2 neutral bosons},\, {\cal CP} = +1 &:& h^0, H^0 \non \\
\mbox{1 neutral boson},\, {\cal CP} = -1  &:& A^0 \non \\
\mbox{2 charged bosons}                   &:& H^+, H^- \non \\
\mbox{3 unphysical Goldstone bosons}      &:& G^0, G^+, G^- .
\EEA
At the tree level, the Higgs boson masses expressed
through $\MZ, \MW$ and $\MA$ are given by
\BEA
\label{mlh}
\mh^2 &=& \edz \KKL \MA^2 + \MZ^2 - 
          \sqrt{(\MA^2 + \MZ^2)^2 - 4 \MA^2\MZ^2 \CQZb} \KKR \\
\label{mhh}
\mH^2 &=& \edz \KKL \MA^2 + \MZ^2 + 
          \sqrt{(\MA^2 + \MZ^2)^2 - 4 \MA^2\MZ^2 \CQZb} \KKR \\
\label{mhp}
\mHp^2 &=& \MA^2 + \MW^2 \\
\label{mg0}
\mG^2 &=& \MZ^2 \\
\label{mgp}
\mGp^2 &=& \MW^2 ,
\EEA
where the last two relations, which assign mass parameters to the
unphysical scalars $G^0$ and $G^{\pm}$, are to be understood in the
Feynman gauge.


\subsection{Evaluation of the \order{\al_t^2}, \order{\al_t \al_b},
            \order{\al_b^2}  contributions}
\label{subsec:gf2mt4eval}

\begin{figure}[htb!]
\begin{center}
\mbox{
\psfig{figure=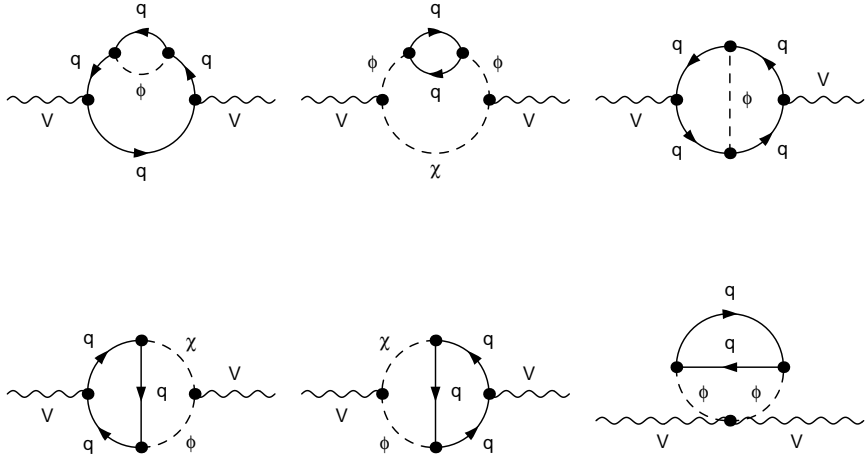,width=8cm,bbllx=140pt,bblly=360pt,
                                          bburx=460pt,bbury=580pt}}
\end{center}
\caption[]{
Generic Feynman diagrams for the vector boson self-energies,\\ 
$(V = \{Z,W\}, q = \{t, b\}, \phi,\chi = \{h, H, A, H^\pm, G, G^\pm\}).$ 
}
\label{fig:fdvb2l}
\end{figure}

In order to calculate the \order{\al_t^2}, \order{\al_t \al_b},
\order{\al_b^2} corrections to $\De\rho$, 
see \refeq{delrho}, the Feynman diagrams generically depicted in
\reffi{fig:fdvb2l} have to be evaluated for the $Z$ boson ($V = Z$)
and the $W$ boson ($V = W$) self-energy. We have taken into account
all possible diagrams involving the $t/b$ doublet and the full Higgs
sector of the MSSM, see \refse{subsec:mssmhiggs}. 

\begin{figure}[htb!]
\vspace{2em}
\begin{center}
\mbox{
\psfig{figure=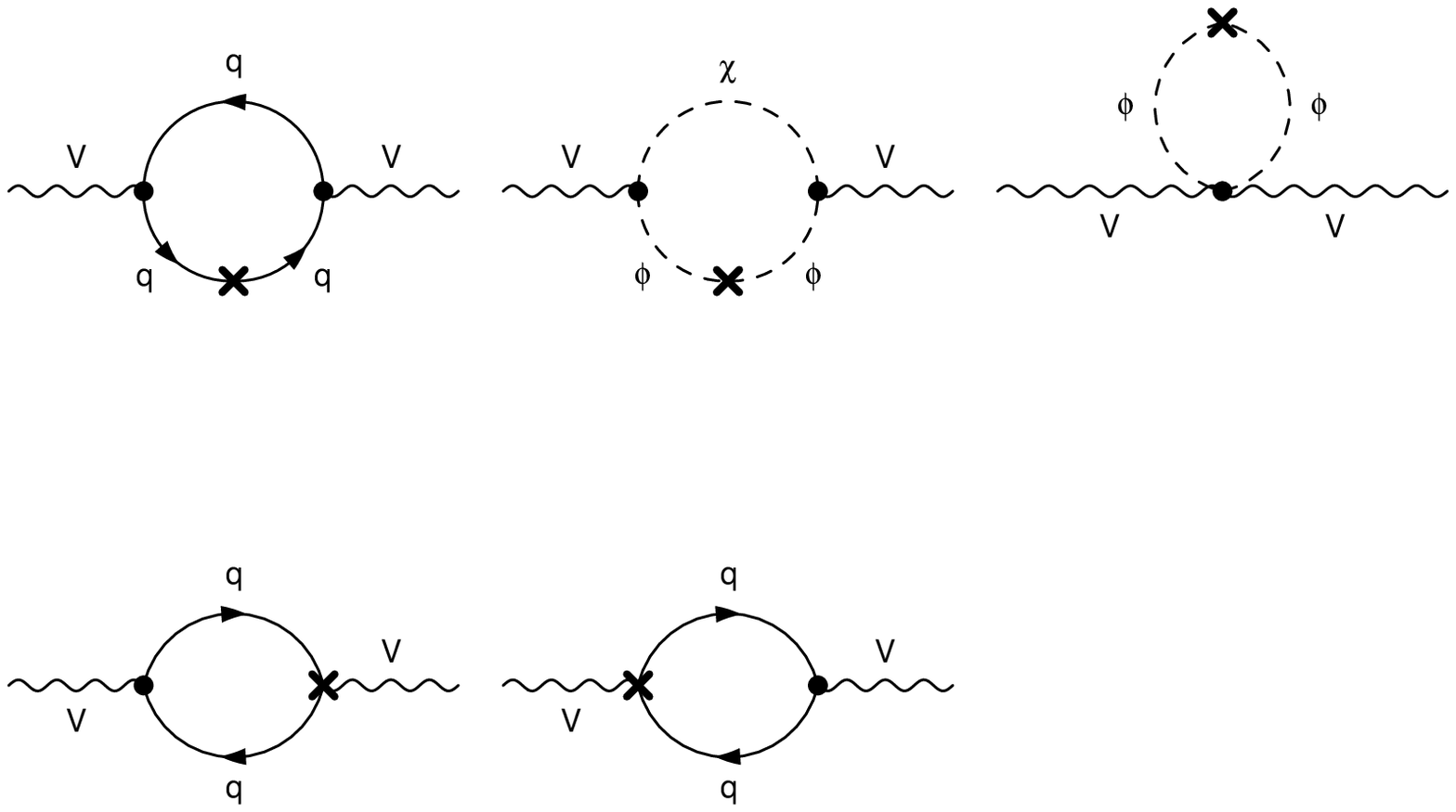,width=8cm,bbllx=130pt,bblly=360pt,
                                          bburx=450pt,bbury=580pt}}
\end{center}
\caption[]{
Generic Feynman diagrams for the vector boson self-energies with
counter term insertion,
$(V = \{Z,W\}, q = \{t, b\}, \phi,\chi = \{h, H, A, H^\pm, G, G^\pm\}).$ 
}
\label{fig:fdvb1lct}
\end{figure}

\begin{figure}[htb!]
\vspace{4em}
\begin{center}
\mbox{
\psfig{figure=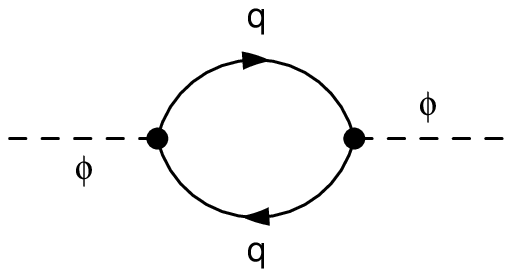,width=3cm,bbllx=100pt,bblly=460pt,
                                         bburx=220pt,bbury=550pt}}
\hspace{3em}
\mbox{
\psfig{figure=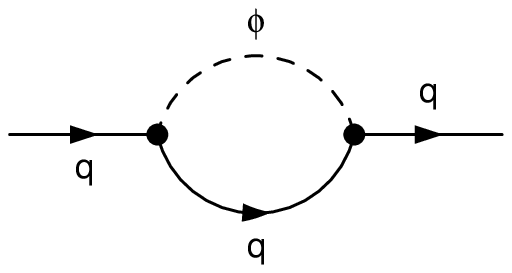,width=3cm,bbllx=100pt,bblly=460pt,
                                         bburx=220pt,bbury=550pt}}
\end{center}
\vspace{-3em}
\caption[]{
Generic Feynman diagrams for the counter term insertions, \\
$(q = \{t, b\}, \phi = \{h, H, A, H^\pm, G, G^\pm\}).$ 
}
\label{fig:fdcti}
\vspace{2em}
\end{figure}

The \twol\ diagrams shown in \reffi{fig:fdvb2l} have to be
supplemented with the corresponding \onel\ diagrams with subloop
renormalization, depicted generically in \reffi{fig:fdvb1lct}. 
The counterterms that enter the calculation are the top mass counter
term, $\de \mt$, the Higgs boson mass counter term, $\de \MA^2$, and
the tadpole counter terms, $\de T_h$ and $\de T_H$. The
renormalization constants have been derived in the on-shell
scheme as outlined in \citere{mhiggslong}. 
The wave function
renormalization constants, entering via the diagrams in
\reffi{fig:fdvb1lct}, drop out as required.  
The Feynman diagrams for the insertions of the fermion and Higgs mass
counter terms are shown in \reffi{fig:fdcti}. 

The amplitudes of all Feynman diagrams, shown in 
\reffis{fig:fdvb2l}--\ref{fig:fdcti}, have been created with the
program {\em FeynArts3}~\cite{feynarts}, making use of the 
MSSM model file~\cite{famssm} 
(where only the non-SM like counter terms had to be added).
The reduction to scalar integrals has been performed with the program
\tc, based on the reduction method of \citere{2lred}. 
As a result we
obtained the analytical expression for $\De\rho$ depending on the
\onel\ functions $A_0$ and $B_0$~\cite{a0b0c0d0} and on the \twol\
function $T_{134}$~\cite{2lred,t134}. For the further evaluation the
analytical expressions for $A_0$, $B_0$ and $T_{134}$ have been
inserted.

In order to derive the leading contributions of
\order{\al_t^2}, \order{\al_t \al_b} and \order{\al_b^2} we extracted the
contributions proportional to $y_t^2$, $y_t y_b$ and $y_b^2$, where
\BE
y_t = \frac{\wz \, \mt}{v \, \Sb}, \quad
y_b = \frac{\wz \, \mb}{v \, \Cb} ~.
\label{ytyb}
\EE
The coefficients of these terms could then be evaluated in the gauge-less 
limit, i.e.\ for $\MW, \MZ \to 0$ (keeping $\cw = \MW/\MZ$ fixed).

For the Higgs masses appearing in the two-loop diagrams we use the
following relations, arising from the gauge-less limit
\BEA
\mHp^2 &=& \MA^2~, \non \\
\mG^2 &=& 0 ~, \non \\
\mGp^2 &=& 0 ~.
\label{chargedHiggsMW0}
\EEA
Applying the corresponding limit also in the neutral $\cp$-even Higgs
sector would yield for the lightest $\cp$-even Higgs-boson mass
$\mh^2 = 0$ (and furthermore $\mH^2 = \MA^2$, $\Sa = -\Cb$, $\Ca = \Sb$). 
Since within the SM the limit $\MHSM \to 0$ turned out to be 
only a poor approximation of the result for arbitrary $\MHSM$, we keep in
our calculation a non-zero $\mh^2$ (which formally is a higher-order
effect). Keeping $\mh$ as a free parameter is also relevant in view of
the fact that the lightest MSSM Higgs boson receives large higher order
corrections~\cite{mhiggs1l},
which shift its upper bound up to $135 \gev$ (for $\msusy \le 1 \tev$
and $\mt = 175 \gev$)~\cite{mhiggslong,mhiggsAEC}. These corrections can
easily be taken into account in this way (in the Higgs contributions at 
one-loop order, however, the tree-level value of $\mh$ should be used).
Keeping $\al$ arbitrary is necessary in order to incorporate non SM-like 
couplings of the lightest $\cp$-even Higgs boson to fermions and gauge
bosons. 

On the other hand, keeping all Higgs-sector parameters
completely arbitrary is not possible, as the underlying symmetry of the
MSSM Lagrangian has to be exploited in order to ensure the
UV-finiteness of the 
two-loop corrections to $\De \rho$. We thus have enforced only those
symmetry relations in the neutral $\cp$-even Higgs sector which are
explicitly needed in order to obtain a complete cancellation of the
UV-divergences.

\bigskip
In the following, we separately consider the \order{\al_t^2}
corrections, corresponding to the limit where $y_b = 0$, and the full
\order{\al_t^2}, \order{\al_t \al_b}, \order{\al_b^2} contributions.
The \order{\al_t^2} corrections are by far the dominant subset within
the SM, i.e.\ the \order{\al_t \al_b} and \order{\al_b^2} corrections
can safely be neglected within the SM. The same is true within the MSSM
for not too large values of $\tan\beta$. Thus, we first consider the
scenario where only the \order{\al_t^2} corrections need to be taken
into account 
and then discuss the result in the case where the \order{\al_t \al_b}
and \order{\al_b^2} corrections are non-negligible.

In the case of the \order{\al_t^2} corrections, no further relations
in the neutral $\cp$-even Higgs sector are necessary, i.e.\ we keep 
the parameters $\mh, \mH$ and $\al$ arbitrary in the evaluation of the
\order{\al_t^2} corrections.
For these contributions also the top Yukawa coupling $y_t$ can be
treated as a free parameter, i.e.\ it is not necessary to use
\refeq{ytyb}.  
As a consistency check of our method we recalculated the corresponding
SM result~\cite{drSMgf2mt4} and found perfect agreement.

\smallskip
Concerning the corrections to $\De\rho$ with $y_b \neq 0$, the
$SU(2)$~structure of the fermion doublet requires further symmetry
relations. Within the Higgs boson sector it is necessary, besides
using \refeq{chargedHiggsMW0}, also to use the relations for the
heavy $\cp$-even Higgs boson mass and the Higgs
mixing angle,
\vspace{-1em}
\BEA
\mH^2 &=& \MA^2 ~, \non \\
\Sa &=& -\Cb ~, \non \\
\Ca &=& \Sb~.
\label{heavyHiggsMW0}
\EEA
On the other hand, $\mh$ can be kept as a
free parameter. The couplings of the lightest $\cp$-even Higgs boson
to gauge bosons and SM fermions,
however, become SM-like, once the mixing angle relations,
\refeq{heavyHiggsMW0}, are used.
Furthermore, the Yukawa couplings can no longer be treated as free
parameters, i.e.\ \refeq{ytyb} has to be employed, which ensures that
the Higgs mechanism governs the Yukawa couplings. 
Also in this case we have evaluated the corresponding SM
corrections. As expected, the corrections arising from 
$y_b \neq 0$ are numerically insignificant within the SM.


\section{The analytical result for the \boldmath{\order{\al_t^2}}
  contributions for the special case \boldmath{$\mh = 0$}}
\label{sec:resmh0}

For illustration, we discuss in this section the result for the
\order{\al_t^2} contributions for the special case where $\mh = 0$. The 
result for the \order{\al_t^2} corrections for arbitrary parameters in
the $\cp$-even Higgs sector can be found in Appendix~A.1, while the 
result for the full \order{\al_t^2}, \order{\al_t\al_b} and
\order{\al_b^2} corrections can be found in Appendix~A.2.
The full results have been included into the code \fh~\cite{feynhiggs}.

\smallskip
In order to simplify the expression for the \order{\al_t^2}
contributions as far as possible, we use in this example 
the relations \refeqs{chargedHiggsMW0} 
and (\ref{heavyHiggsMW0}) as well as $\mh = 0$.
The only remaining parameters in this case 
are the top quark mass, $\mt$, the $\cp$-odd
Higgs boson mass, $\MA$, and $\tb$ (or $\sbe \equiv \tb/\sqrt{1 + \TQb}$). 

The analytical result obtained as described in
\refse{subsec:gf2mt4eval} can conveniently be expressed in terms of 
\BE
\label{defa}
A \equiv \frac{\mt^2}{\MA^2} .
\EE
\newcommand{\sqa}{\sqrt{1 - 4\,A}}

\noindent
The \twol\ contribution to the \rp\ then reads:
\BEA
\De\rho_{1,\hi,\mh=0}^{\SU} &=& 3 \, \frac{\gf^2}{128 \,\pi^4} \, \mt^4 \, 
                      \frac{1 - \sbe^2}{\sbe^2 \, A^2} \times \non \\
&& \Bigg\{ 
   \Liz{ \KL 1 - \sqa \KR/2} \frac{8}{\sqa} \La \non \\
&& - 2\, \Liz{1 - \ed{A}} \KKL 5 - 14 A + 6 A^2 \KKR \non \\
&& + \log^2(A) \KKL 1 + \frac{2}{\sqa} \La \KKR \;
   - \log(A) \Big[ 2 - 20 \, A \Big] \non \\
&& - \log^2 \KL \frac{1 - \sqa}{2} \KR \frac{4}{\sqa} \La \non \\ 
&& + \log \KL \frac{1 - \sqa}{1 + \sqa} \KR 
     \sqa (1 - 2 \, A) \non \\
&& - \log\KL |1/A - 1| \KR \, (A - 1)^2 \non \\
&& + \pi^2 \KKL \frac{2\sqa}{-3 + 12 \, A} \La
                + \ed{3} - 2\, A^2 \frac{\sbe^2}{1 - \sbe^2} \KKR 
   - 17 A + 19 \frac{A^2}{1 - \sbe^2} 
   \Bigg\} ~,
\label{drallma}
\EEA
with 
\BE
\La = 3 - 13\,A + 11\,A^2 .
\EE
In the limit of large $\tb$ (i.e.\ $(1 - \sbe^2) \ll 1$) one obtains
\BE
\De\rho_{1,\hi,\mh=0}^{\SU} = 3 \, \frac{\gf^2}{128 \,\pi^4} \, \mt^4 \, 
\KKL \frac{19}{\sbe^2} - 2\,\pi^2 + \orderm{1 - \sbe^2} \KKR .
\label{drallmalargetb}
\EE
Thus for large $\tb$ the SM limit with $\MH^{\SM} \to 0$ is reached.


\smallskip
In order to investigate the decoupling behavior of 
$\De\rho_{1,\hi,\mh=0}^{\SU}$, 
the result for $\De\rho_{1,\hi,\mh=0}^{\SU}$ in \refeq{drallma}
can be expanded for small values of
$A$, i.e.\ for large values of $\MA$:
\BEA
\De\rho_{1,\hi,\mh=0}^{\SU} &=& 3 \, \frac{\gf^2}{128 \,\pi^4} \, \mt^4 \, 
                      \times \non \\
&& \Bigg\{
   19 - 2 \pi^2 \non \\
&& -\frac{1 - \sbe^2}{\sbe^2} \Bigg[ \KL \log^2 A + \frac{\pi^2}{3} \KR
           \KL 8 A + 32 A^2 + 132 A^3 + 532 A^4 \KR \non \\
\label{drlargema}
&& + \log(A) \ed{30} \KL 560 A + 2825 A^2 + 11394 A^3 + 45072 A^4 \KR 
     \\
&& - \ed{1800} \KL 2800 A + 66025 A^2 + 300438 A^3 + 1265984 A^4 \KR
   + {\cal O}\KL A^5 \KR \Bigg] 
   \Bigg\}~. \non
\EEA
In the limit $A \to 0$ one obtains
\BE
\De\rho_{1,\hi,\mh=0}^{\SU} = 3 \, \frac{\gf^2}{128 \,\pi^4} \, \mt^4 \, 
 \KKL 19 - 2\,\pi^2 \KKR + \cO(A) ,
\EE
i.e.\ exactly the SM limit for $\MH^{\SM} \to 0$ is
reached. 
This constitutes a consistency check, since in the limit $A \to 0$ 
the heavy Higgs bosons are decoupled from the theory. Thus only the
lightest $\cp$-even Higgs boson should remain, which in the
\order{\al_t^2} approximation (neglecting higher-order corrections)
has the mass $\mh = 0$. 

An expansion for small values of $\MA$ as well as an analysis of the
quality of these expansions can be found in \citere{drMSSMgf2}.

The more general expressions, i.e.\ with $\mh \neq 0$, at
\order{\al_t^2}, \order{\al_t \al_b}, and \order{\al_b^2} can be found
in the appendix.


\section{Numerical analysis}
\label{sec:numanal}

\subsection{The \order{\al_t^2} contributions}
\subsubsection{Comparison for $\De\rho$}
\label{subsec:delrho}

\begin{figure}[htb!]
\vspace{1em}
\begin{center}
\mbox{
\epsfig{figure=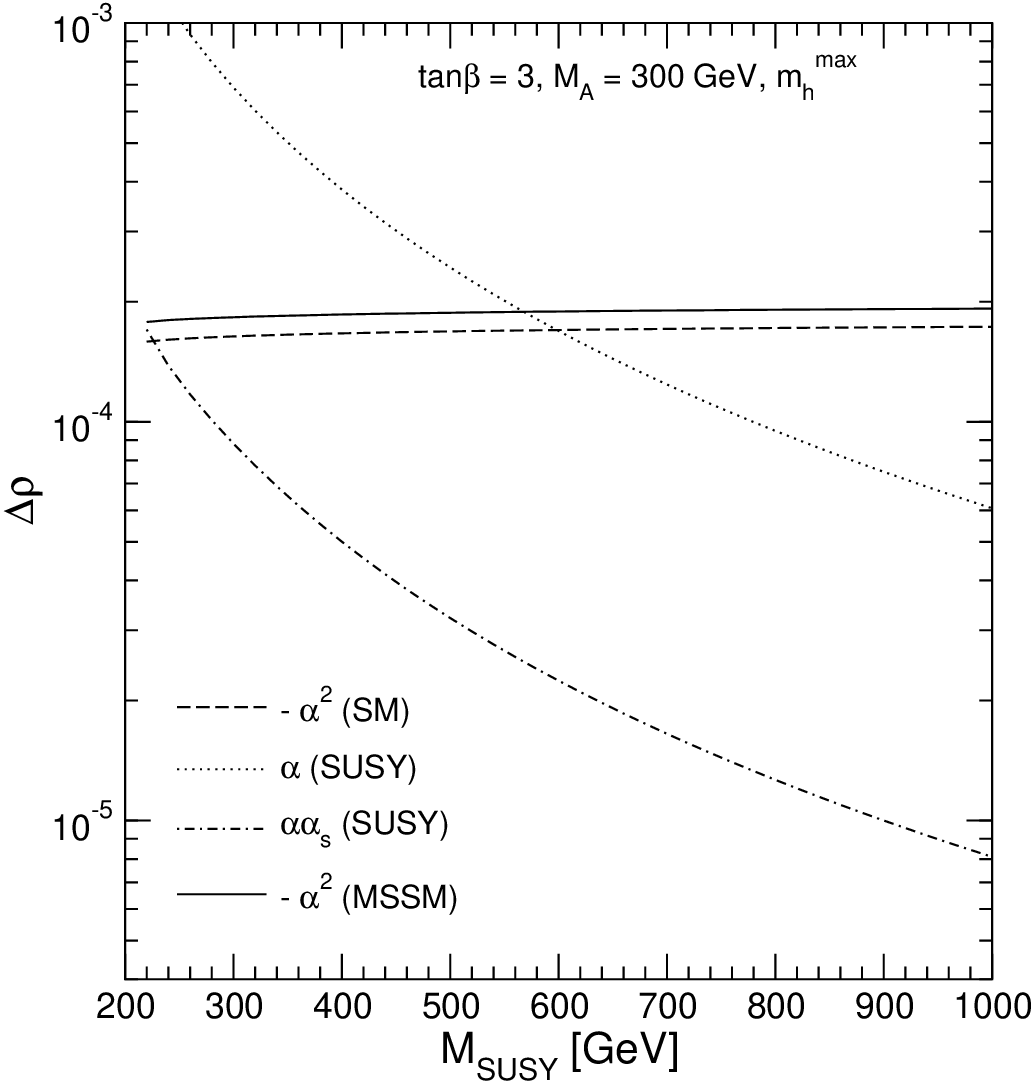,width=7cm,height=6.5cm} 
\hspace{1em}
\epsfig{figure=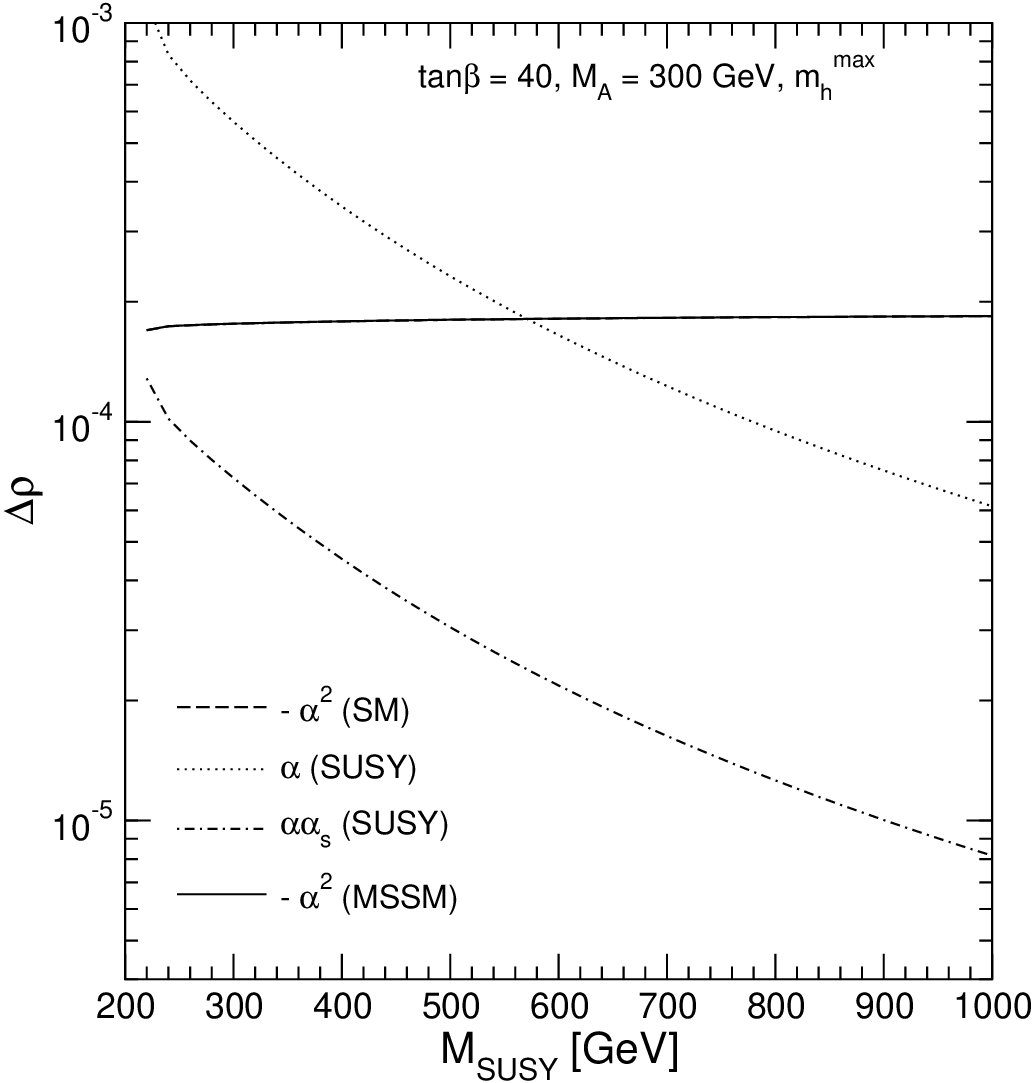,width=7cm,height=6.5cm} 
}
\end{center}
\caption[]{
The contribution of the leading \order{\al_t^2} MSSM corrections,
$\De\rho_{1,\hi}^{\SU}$, \refeq{drfull}, is shown as a function of
$\msusy$ for $\MA = 300 \gev$ and 
$\tb = 3$ (left plot) or $\tb = 40$ (right plot) in the $\mhmax$
scenario. 
$\De\rho_{1,\hi}^{\SU}$ is compared with the leading
\order{\al_t^2} SM contribution and with the
leading MSSM corrections originating from the $\Stop/\Sbot$ sector of
\order{\al} and \order{\al\als}.
Both \order{\al_t^2} contributions are
negative and are for comparison shown with reversed sign.
In the right plot the \order{\al_t^2} corrections differ by about
$1.5 \times 10^{-7}$, which is not visible in the plot. 
}
\label{fig:delrho_Msusy}
\end{figure}

In \reffi{fig:delrho_Msusy} the size of the leading \order{\al_t^2}
MSSM corrections, 
\refeq{drfull}, is compared with the leading 
\order{\al_t^2} contribution
in the SM~\cite{drSMgf2mt4}, with the leading MSSM
corrections arising from the $\Stop/\Sbot$ sector at
\order{\al}~\cite{dr1lB}, and with the corresponding gluon-exchange
contributions of \order{\al\als}~\cite{dr2lA} (the \order{\al\als}
gluino-exchange contributions~\cite{dr2lA}, which go to zero for large
$\mgl$, have been omitted here). 
The numerical effects of the different contributions to $\De\rho$ are
shown as a function of a common SUSY mass scale, $\msusy$ (which
enters the diagonal entries in the $\Stop$~mass matrix). For the
MSSM parameters we have chosen the values as specified in the
$\mhmax$~benchmark scenario~\cite{LHbenchmark}, i.e.\ 
$\Xt = 2\, \msusy$, where $\mt\Xt$
is the off-diagonal entry in the $\Stop$~mass matrix. For our
conventions in the $\Stop$~sector, see \citere{mhiggslong}. The other
parameters are  
$\mu = 200 \gev, \Ab = \At$. $\mu$ is the Higgs mixing parameter
and $A_{t,b}$ are the trilinear Higgs-$\Stop,\Sbot$ couplings, 
respectively. While $\MA$ has been set to $\MA = 300 \gev$, for $\tb$
we have chosen two typical values, $\tb = 3$ as a low and $\tb = 40$
as a high value. 
(Smaller $\tb$ values within the $\mhmax$~scenario, where $\mt$ is
fixed to $\mt = 174.3 \gev$ and 
$\msusy \le 1000 \gev$, are disfavored by the LEP 
Higgs boson searches~\cite{lephiggs,tbexcl}.)
From these
parameters the values for $\mh$, $\mH$ and $\al$ have been obtained. 
For the numerical evaluation of the $\cp$-even Higgs boson sector, we
have used the results from the $t/\Stop$ sector as presented in
\citeres{mhiggslong,mhiggsletter,mhiggslle}.
The SM \order{\al_t^2} corrections, \refeq{drSMgf2},  have
been evaluated using the result of $\mh$ as the SM Higgs boson
mass. 
$\msusy$ enters the MSSM \order{\al_t^2} corrections (where as
described above, we have neglected the SUSY loop contributions) only
indirectly via its effect on $\mh$.

\reffi{fig:delrho_Msusy} shows the decoupling of the effects of scalar
quark loops with increasing $\msusy$. The $\oaas$ SUSY corrections are
always about an order of magnitude smaller than the $\oa$ squark loop
contributions to $\De\rho$. 
The decoupling with $\msusy$ indicates that for large values of
$\msusy$ the contributions from quarks and scalar quarks within the
MSSM essentially reduce to the quark loop corrections. 
This motivates to approximate the \order{\al_t^2} corrections in
the full MSSM by the Two-Higgs-Doublet model part.

The \order{\al_t^2} corrections 
involving quarks and the Higgs sector of the MSSM turn out to be
larger than the $\oaas$ SUSY corrections for all values of 
$\msusy \gsim 200 \gev$. This is related to the enhancement by the
prefactor $\mt^4/\MW^4$. 
The \order{\al_t^2} corrections even exceed the $\oa$ squark loop
corrections for $\msusy \gsim 600 \gev$, 
i.e.\ these contributions can compensate each other as they enter
with different sign. This applies also for the no-mixing
scenario ($\Xt = 0$, $\msusy = 2000 \gev$)~\cite{LHbenchmark}, which is
not shown here.

In \reffi{fig:delrho_mh} we analyze the dependence of the
\order{\al_t^2} contributions to $\De\rho$ on the lightest $\cp$-even
Higgs boson mass, $\mh$. For the MSSM parameters we have again chosen
values as specified for the $\mhmax$~and the no-mixing scenario. 
While $\tb$ has been fixed to $\tb = 3, 40$, the $\cp$-odd Higgs boson
mass has been varied from $50 \gev$ to $1000 \gev$. 

As can be seen in \reffi{fig:delrho_mh}, the \order{\al_t^2} MSSM
contribution is of \order{10^{-4}}. It is always larger than the
corresponding SM result. In the limit of large $\MA$, i.e.\
at the endpoint of the $\mh$ spectrum, the difference of the SM and the
MSSM result are numerically 
very small. This is in accordance with the decoupling behavior that
we have discussed analytically for the special case with $\mh = 0$,
see \refse{sec:resmh0}.

%
\begin{figure}[htb!]
\vspace{3em}
\begin{center}
\mbox{
\epsfig{figure=delrhoMT2Yukfull02.bw.eps,width=7cm,height=6.5cm} 
\hspace{1em}
\epsfig{figure=delrhoMT2Yukfull06.bw.eps,width=7cm,height=6.5cm} 
}
\end{center}
\caption[]{
The contribution of the leading \order{\al_t^2} MSSM corrections,
$\De\rho_{1,\hi}^{\SU}$, is shown as a function of $\mh$ for
$\tb = 3$ (left plot) and for $\tb = 40$ (right plot) in the $\mhmax$
and the no-mixing scenario.
$\De\rho_{1,\hi}^{\SU}$ is compared with the leading
\order{\al_t^2} SM contribution.
}
\label{fig:delrho_mh}
\end{figure}

\begin{figure}[htb!]
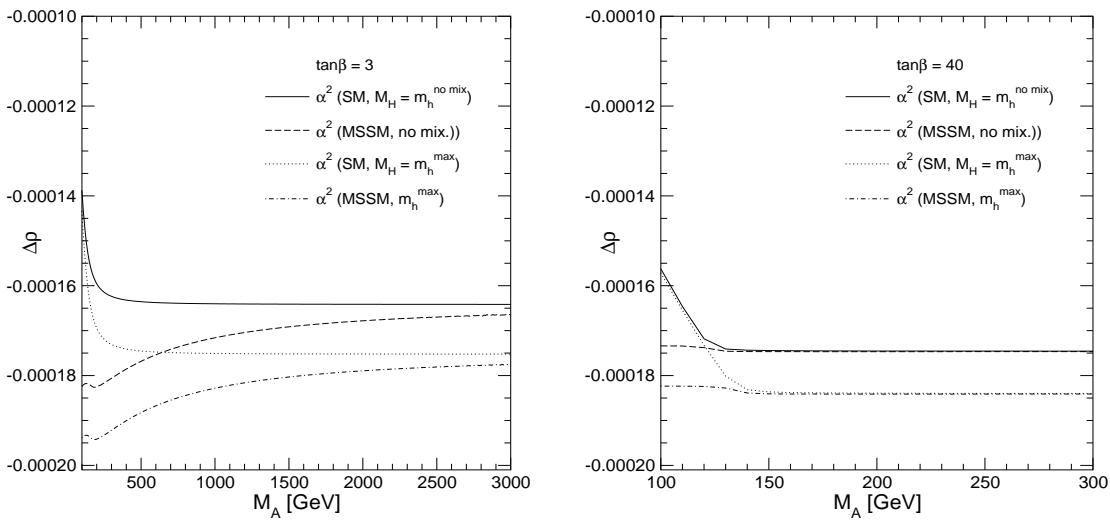

\vspace{1em}
\begin{center}
\mbox{
\epsfig{figure=delrhoMT2Yukfull12b.bw.eps,width=7cm,height=6.8cm} 
\hspace{1em}
\epsfig{figure=delrhoMT2Yukfull16.bw.eps,width=7cm,height=6.8cm} 
}
\end{center}
\caption[]{
The \order{\al_t^2} MSSM contribution to $\De\rho$ in the $\mhmax$ and
the no-mixing scenario is compared with the corresponding SM result with
$\MHSM = \mh$. $\mh$ is obtained in the left (right)
plot from varying $\MA$ from  $100 \gev$ to $3000 (300) \gev$, 
while keeping $\tb$ fixed at $\tb = 3 (40)$.
}
\label{fig:madep}
\end{figure}

In \reffi{fig:madep} the decoupling behavior for large $\MA$ of 
$\De\rho^{\SU}_{1,{\rm Higgs}}$ is analyzed numerically. 
Since $\MA$ is the only non-SM scale that directly enters 
$\De\rho^{\SU}_{1,{\rm Higgs}}$, the result should become SM-like once
$\MA$ is taken to very large values. This is shown in the $\mhmax$ and
the no-mixing scenario for $\tb$ fixed to $\tb = 3 (40)$ in the left
(right) plot of \reffi{fig:madep}. The SM value of 
$\De\rho^{\SM}_{1,{\rm Higgs}}$ is obtained for $\MHSM = \mh$.
While for the small $\tb$ value the decoupling is very slow 
and is reached only for $\MA \gsim 3 \tev$, for the large
$\tb$ value $\De\rho^{\SU}_{1,{\rm Higgs}}$ decouples already for very
small $\MA$ values. This numerical behavior is analogous to the
analytical result described in \refse{sec:resmh0} for the 
$\mh \to 0$ limit. 
However, it should also be noted that for small values of $\MA$ the
behavior of the SM and the MSSM contributions is very different. 
While the SM contribution depends sensitively on $\MHSM$, in the MSSM
for $\MA \gsim 100 \gev$ (corresponding to $\mh \gsim 90 \gev$, see
\reffi{fig:delrho_mh}) the dependence on the Higgs boson masses is
much less pronounced, see in particular the right plot of
\reffi{fig:madep}.


\subsubsection{Effects on precision observables}
\label{subsec:precobs}

In this section the numerical effect of the \order{\al_t^2}
corrections on the electroweak precision observables, $\MW$ and
$\sweff$ is analyzed. In addition to the MSSM 
\order{\al_t^2} correction to $\de\MW$ and $\de\sweff$, we also present
the effective change from the SM result (where the SM Higgs boson mass
has been set to $\mh$) to our new MSSM result.

\begin{figure}[htb!]
\begin{center}
\mbox{
\epsfig{figure=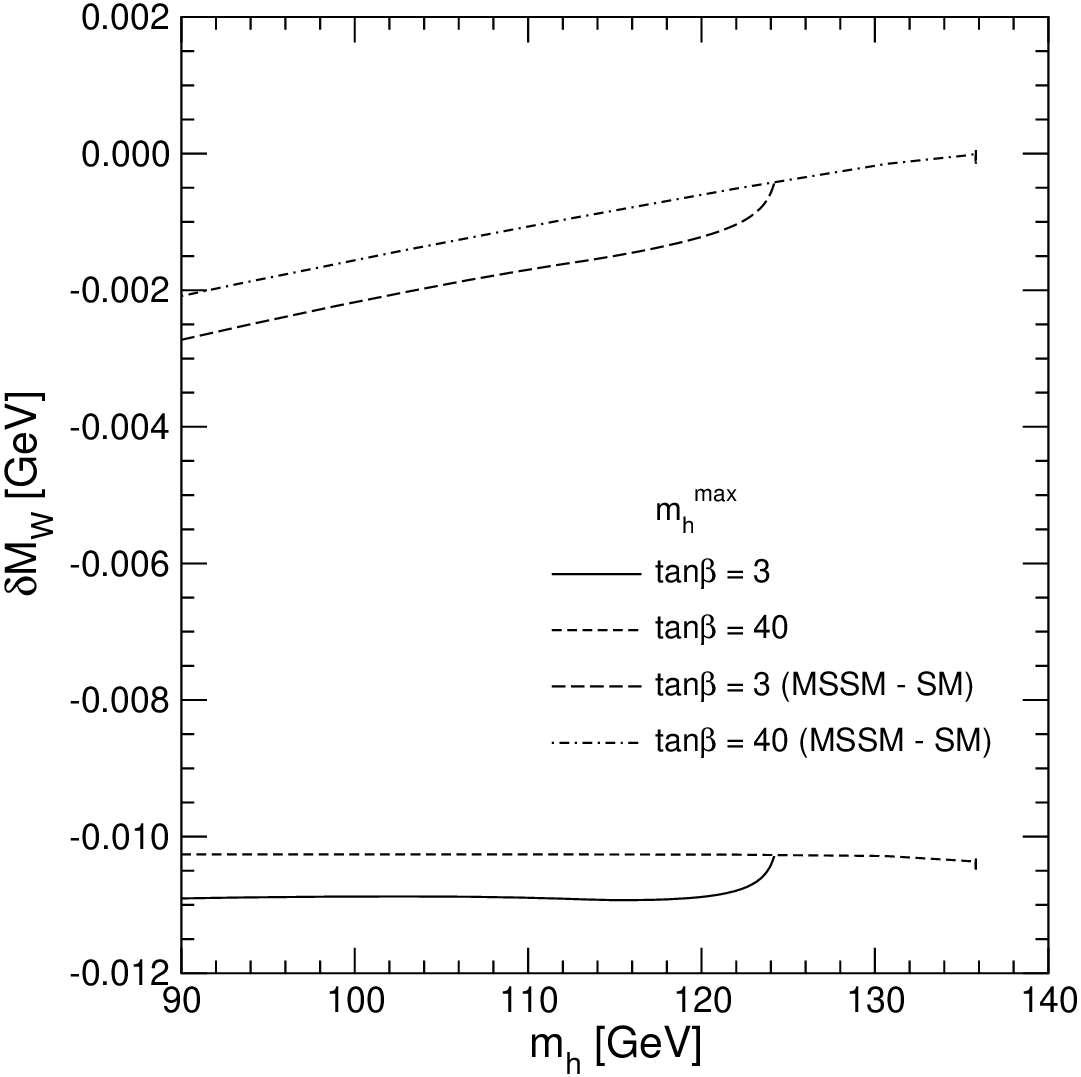,width=7cm,height=6.8cm} 
\hspace{1em}
\epsfig{figure=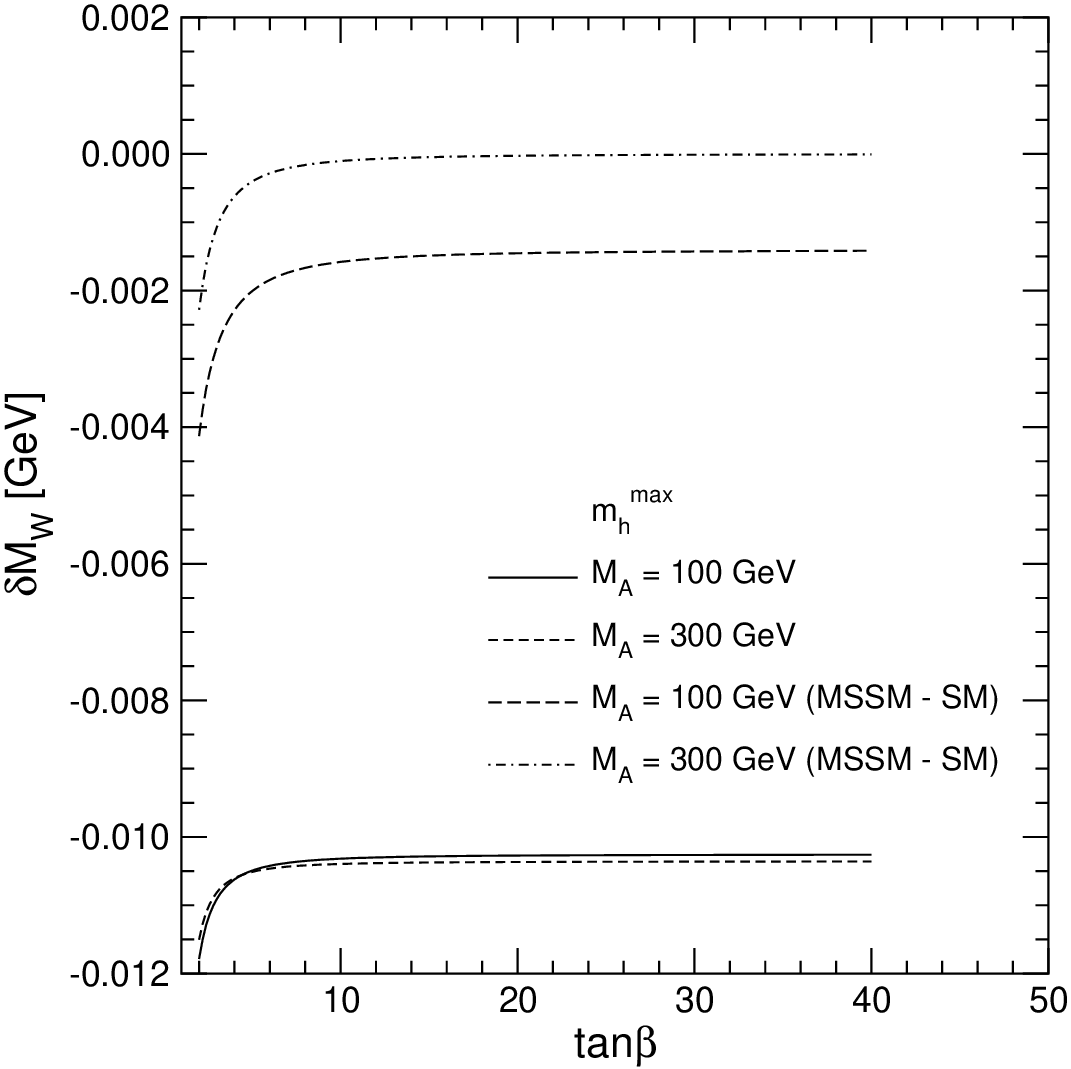,width=7cm,height=6.8cm} 
}
\end{center}
\caption[]{
The absolute \order{\al_t^2} MSSM contribution and the effective change 
in $\de\MW$ is shown for $\msusy = 1000 \gev$ in
the $\mhmax$~scenario. 
The other parameters are $\mu = 200 \gev, \Ab = \At$. $\mh$ is
obtained in the left (right) plot from varying $\MA$ from 
$50 \gev$ to $1000 \gev$, while keeping $\tb$ fixed at $\tb = 3, 40$
(from varying $\tb$ from 2 to 40, while keeping $\MA$ fixed at
$\MA = 100, 300 \gev$.)
}
\label{fig:delMW}
\end{figure}
%
\begin{figure}[htb!]
\vspace{2em}
\begin{center}
\mbox{
\epsfig{figure=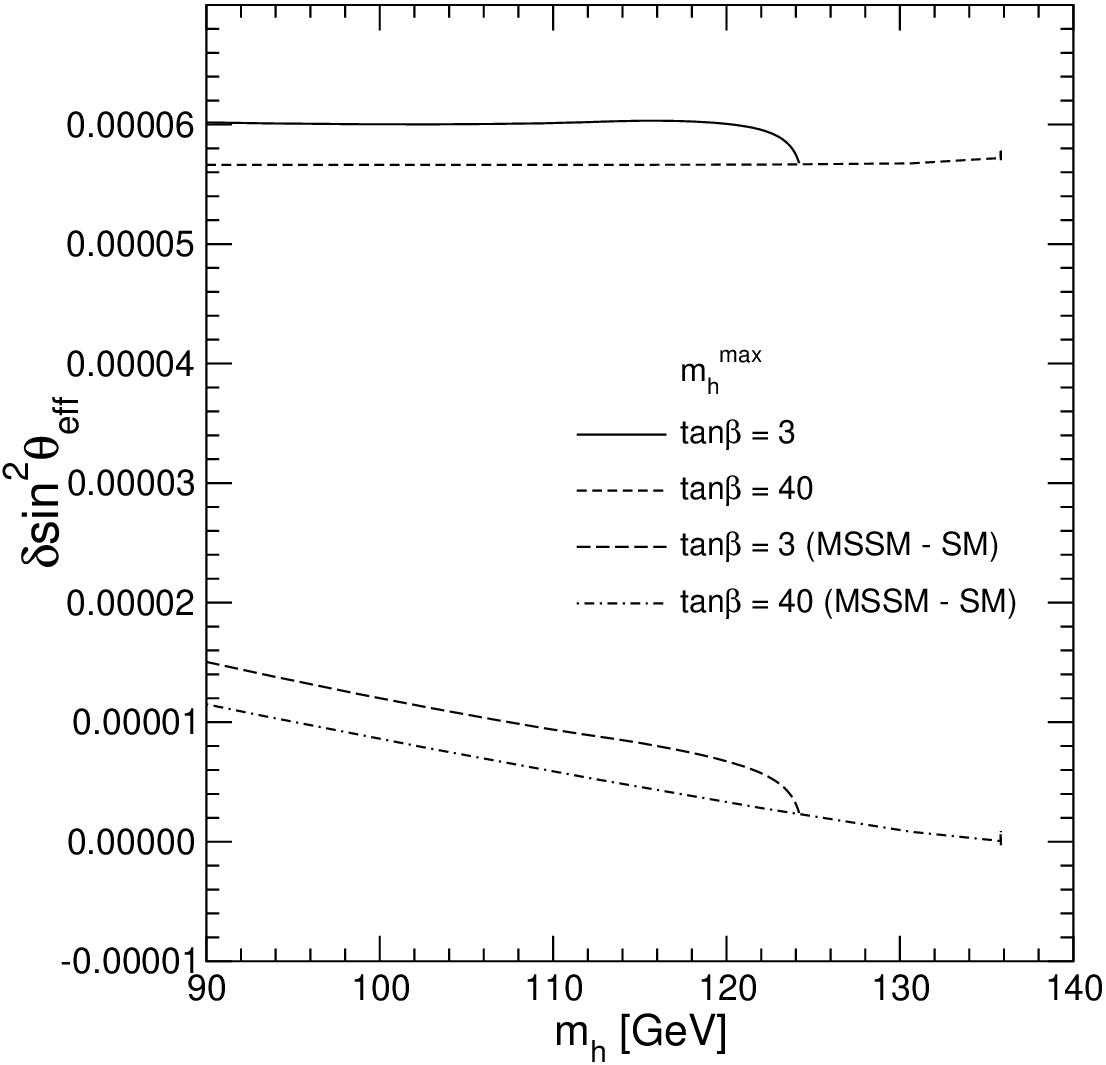,width=7cm,height=6.8cm} 
\hspace{1em}
\epsfig{figure=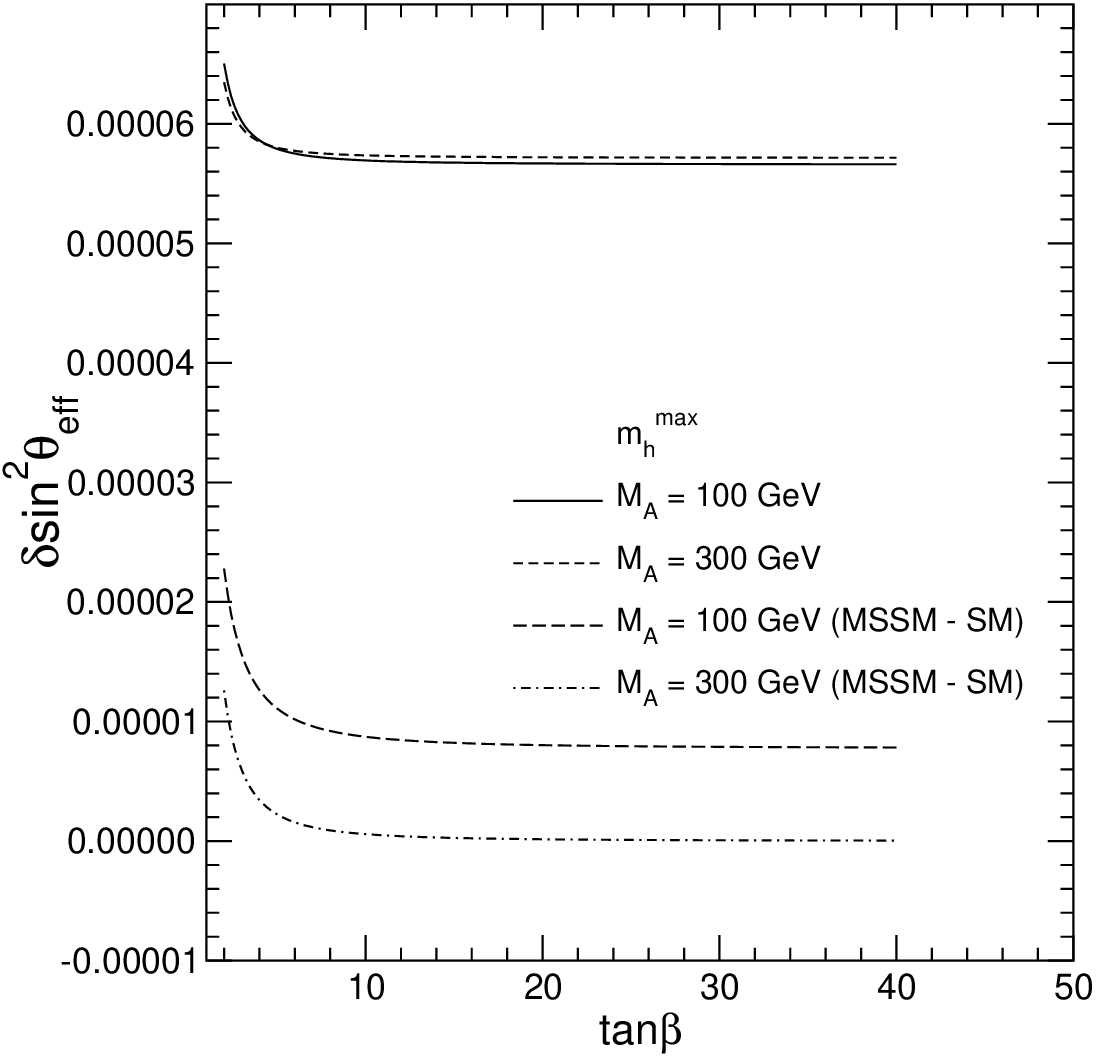,width=7cm,height=6.8cm} 
}
\end{center}
\caption[]{
The absolute \order{\al_t^2} MSSM contribution and the effective change 
in $\de\sweff$ is shown for $\msusy = 1000 \gev$ in
the $\mhmax$~scenario. 
The other parameters are $\mu = 200 \gev, \Ab = \At$. $\mh$ is
obtained in the left (right) plot from varying $\MA$ from 
$50 \gev$ to $1000 \gev$, while keeping $\tb$ fixed at $\tb = 3, 40$
(from varying $\tb$ from 2 to 40, while keeping $\MA$ fixed at
$\MA = 100, 300 \gev$.)
}
\label{fig:delsw2eff}
\end{figure}

In \reffi{fig:delMW} the absolute contribution and the 
effective change for the $W$~boson mass is
presented. For the numerical evaluation we have chosen the $\mhmax$
benchmark scenario (where $\msusy = 1000 \gev$). In the left plot $\tb$ is
fixed to $\tb = 3, 40$, while $\MA$ is varied from $50 \gev$ to 
$1000 \gev$, resulting in the Higgs boson mass $\mh$. 
The effect of the \order{\al_t^2} MSSM contributions on $\de\MW$ 
amounts up to $-12 \mev$. For large $\MA$ and/or large $\tb$ it
saturates at about $-10 \mev$.
The effective change in $\MW$ is significantly smaller. It amounts up
to $-3 \mev$ and goes to zero for large $\MA$ as expected from the
decoupling behavior.
In the right plot of \reffi{fig:delMW} $\de\MW$ is shown as a function
of $\tb$. $\MA$ is kept fixed to $\MA = 100, 300 \gev$.  The effect of
$\De\rho^{\SU}_{1,\hi}$ saturates for large $\tb$. 
For a small $\cp$-odd Higgs boson mass, 
$\MA = 100 \gev$, a shift of $-2 \mev$ in $\MW$ remains also in the
limit of large $\tb$, since the two Higgs doublet sector does not
decouple from the MSSM. For large $\MA$, $\MA = 300 \gev$, for nearly
all $\tb$ values the effective change in $\MW$ is small.

The absolute contribution and the effective change for $\de\sweff$ is
shown in 
\reffi{fig:delsw2eff} for the same parameters as in \reffi{fig:delMW}. 
The absolute effect is around $+6 \times 10^{-5}$. 
The effective change ranges between 
$+3 \times 10^{-5}$ for small $\tb$ and small $\MA$ and 
approximately zero for large $\tb$ and large $\MA$.

The effects of the \order{\al_t^2} MSSM corrections in $\MW$ and
$\sweff$ discussed above are smaller than the current experimental
errors, $\de\MW^{\rm exp} = 34 \mev$ and 
$\de\sweff^{\rm exp} = 17 \times 10^{-5}$~\cite{grueni}.
However, their inclusion is crucial in order to reduce the theoretical
uncertainties from unknown higher order corrections within the MSSM to
a similar level as in the SM of 
$\de\MW^{\rm theo} \approx \pm 5 \mev$ and
$\de\sweff^{\rm theo} \approx \pm 7 \times 10^{-5}$
\cite{deltarferm,EWPOSM}.
Achieving this level of theoretical accuracy will be mandatory in
particular in view of the prospective accuracies at a future linear
collider running on the $Z$~peak and the $WW$~threshold (GigaZ), 
$\de\MW^{\rm exp,fut} \approx 7 \mev$ and 
$\de\sweff^{\rm exp,fut} \approx 1.3 \times 10^{-5}$
\cite{moenig,gigaz,EWPOSM}.


\subsection{The \order{\al_t^2}, \order{\al_t \al_b}, and \order{\al_b^2}
            contributions} 

In this section the numerical effect of the \order{\al_t^2},
\order{\al_t \al_b}, and \order{\al_b^2} corrections on $\De\rho$ is 
analyzed. As discussed in \refse{subsec:gf2mt4eval}, for these
corrections is was necessary to employ the Higgs sector restrictions
as given in \refeqs{chargedHiggsMW0} and (\ref{heavyHiggsMW0}). This
implies that the couplings of the lightest $\cp$-even Higgs boson to
gauge bosons and SM fermions are SM-like. Corrections enhanced by
$\tb$ thus arise only from the heavy Higgs bosons, while the
contribution from the lightest $\cp$-even Higgs boson resembles the SM
one.

\begin{figure}[htb!]
\begin{center}
\mbox{
\epsfig{figure=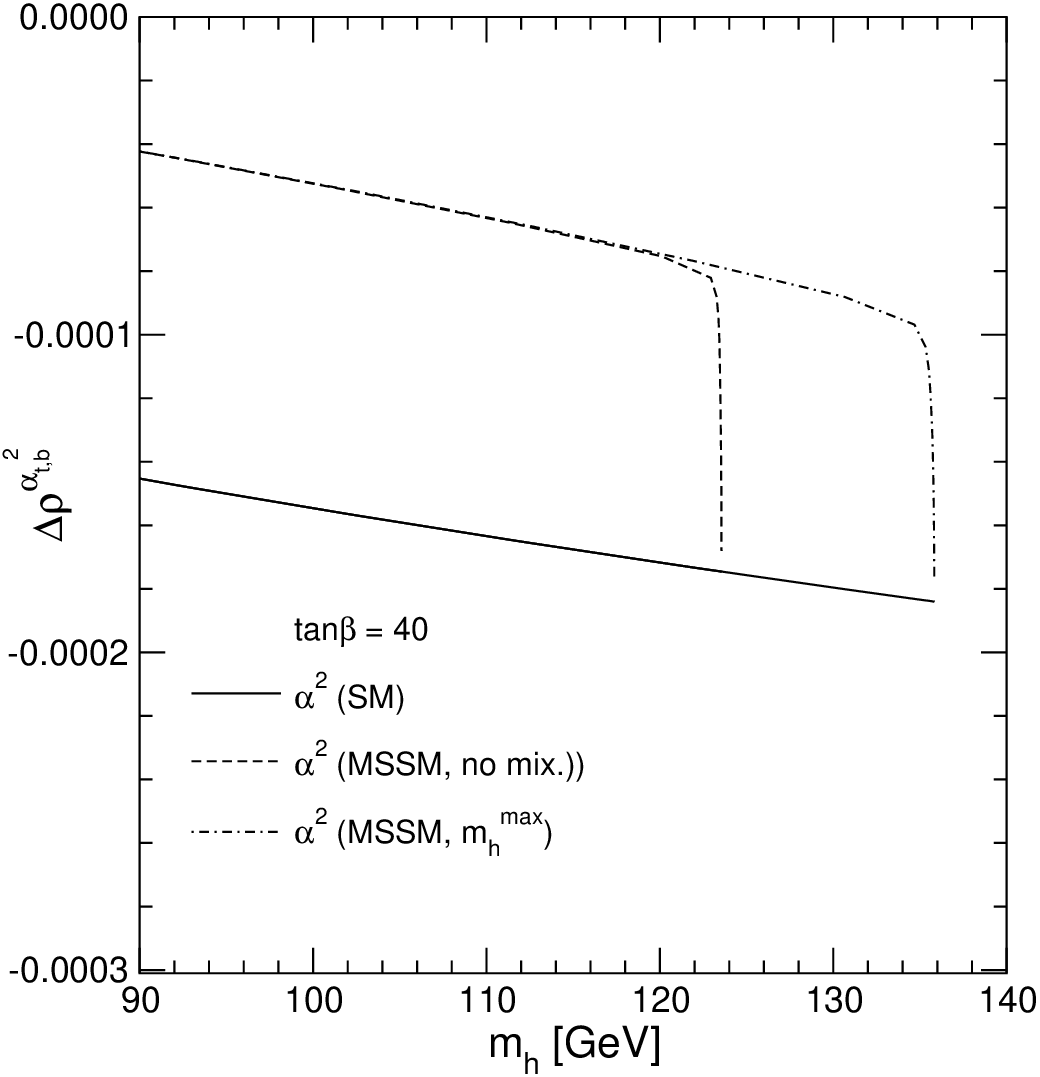,width=7cm,height=6.8cm} 
\hspace{1em}
\epsfig{figure=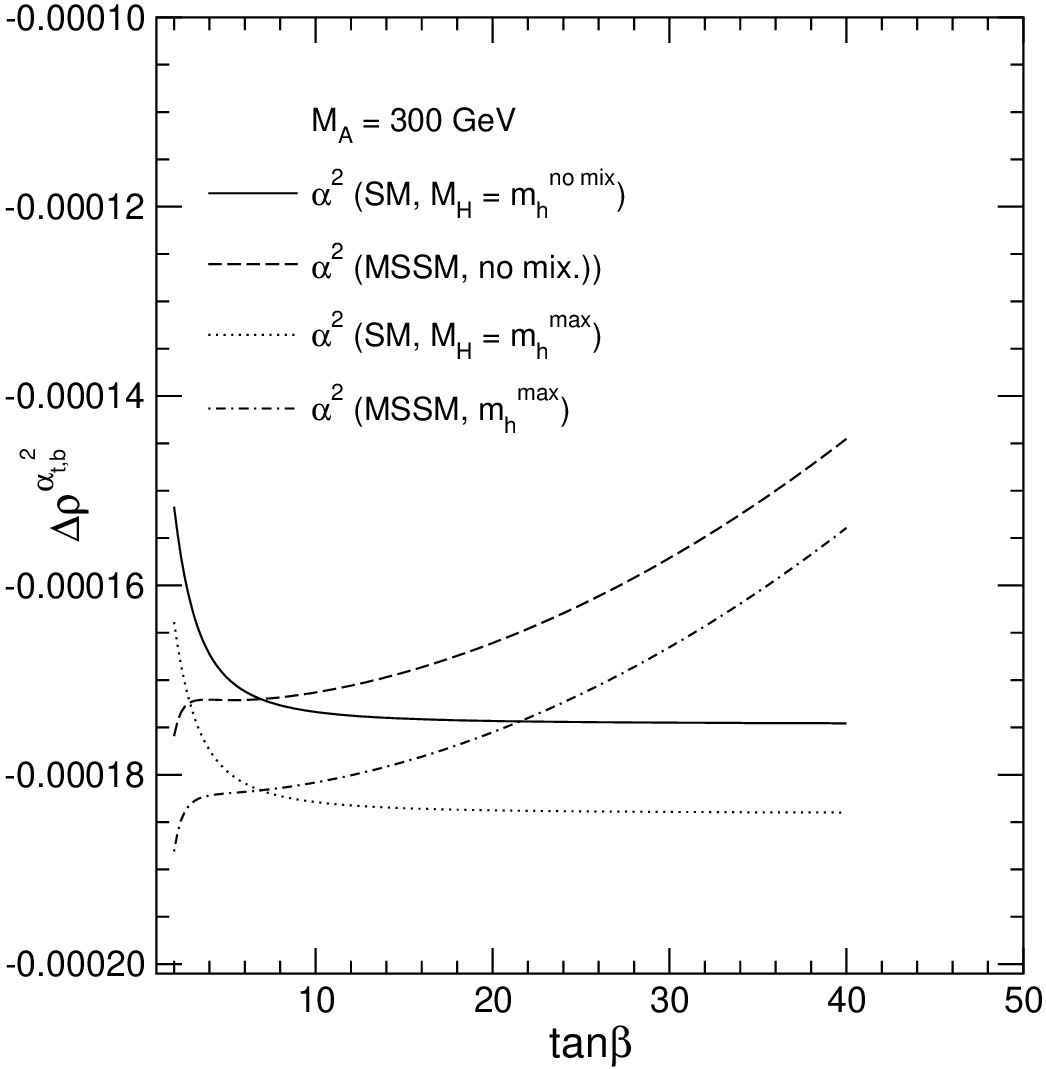,width=7cm,height=6.8cm} 
}
\end{center}
\caption[]{
The \order{\al_t^2}, \order{\al_t \al_b}, and \order{\al_b^2} MSSM
contribution to $\De\rho$ in the $\mhmax$ and the no-mixing scenario
is compared with the corresponding SM result with $\MHSM = \mh$. 
In the left plot $\tb$ is fixed to $\tb = 40$, while $\MA$
is varied from $50 \gev$ to $1000 \gev$. In the right plot $\MA$ is
set to $300 \gev$, while $\tb$ is varied. The bottom quark mass is set
to $\mb = 4.25 \gev$.
}
\label{fig:delrhotb}
\end{figure}

In \reffi{fig:delrhotb} we show the result for the \order{\al_t^2},
\order{\al_t \al_b}, and \order{\al_b^2} MSSM contributions to 
$\De\rho$ in the $\mhmax$ and the no-mixing scenario, 
compared with the corresponding SM result with $\MHSM = \mh$. 
In the left plot $\tb$ is fixed to $\tb = 40$ and $\MA$ is varied from
$50 \gev$ to $1000 \gev$. In the right plot $\MA$
is fixed to $\MA = 300 \gev$, while $\tb$ is varied. 

For large $\tb$ the \order{\al_t \al_b} and \order{\al_b^2}
contributions yield a significant effect from the heavy Higgs bosons
in the loops, entering with the other sign than the \order{\al_t^2}
corrections, while the contribution of the lightest Higgs boson is
SM-like. As one can see in \reffi{fig:delrhotb}, for large $\tb$ the
MSSM contribution to $\De\rho$ is smaller than the SM value. For large
values of $\MA$, the SM result is recovered. 
The effective change in the predictions for the precision observables
from the \order{\al_t \al_b} and \order{\al_b^2} corrections can
exceed the one from the \order{\al_t^2} corrections. It can amount up
to $\de\MW \approx +5 \mev$ and $\de\sweff \approx -3\times 10^{-5}$
for $\tb = 40$.


\section{Constraints on the top Yukawa coupling in the 
SM and the MSSM}

The $\De\rho^{\al_{t,b}^2}_{1, \rHiggs}$ corrections in the SM and the
MSSM are of particular interest, 
since these are the leading corrections in which the top and bottom Yukawa
couplings, i.e.\ the coupling of Higgs bosons to top and bottom quarks,
enter the predictions for the electroweak precision observables. Thus,
the electroweak precision tests of the SM and the MSSM provide some 
sensitivity to the Yukawa couplings in these models.

In order to exemplify this sensitivity, we use a simple approach in
which we treat the top Yukawa coupling in the SM and the MSSM as a
free parameter. While a complete calculation of top and bottom
contributions, as discussed in the previous sections, requires the
relation between the Yukawa coupling and fermion mass within the SM
and the MSSM, this relation is not formally needed if one restricts to
the top contributions only. Numerically, this is an excellent
approximation within the SM and also in the MSSM for not too
large~$\tb$. 

In the following we analyze the sensitivity to the top Yukawa coupling in
the SM and the MSSM. Since in the MSSM contributions beyond the
$\De\rho^{\al_{t}^2}_{1, \rHiggs}$ corrections are not yet known, for
this comparison we restrict the SM contributions also to the leading 
electroweak $\De\rho^{\al_{t}^2}_{1, \rHiggs}$ term~\cite{drSMgf2mt4},
neglecting the formally subleading electroweak two-loop corrections
to the precision observables~\cite{deltarferm}, which can, however, be
of similar size.

\begin{figure}[htb!]
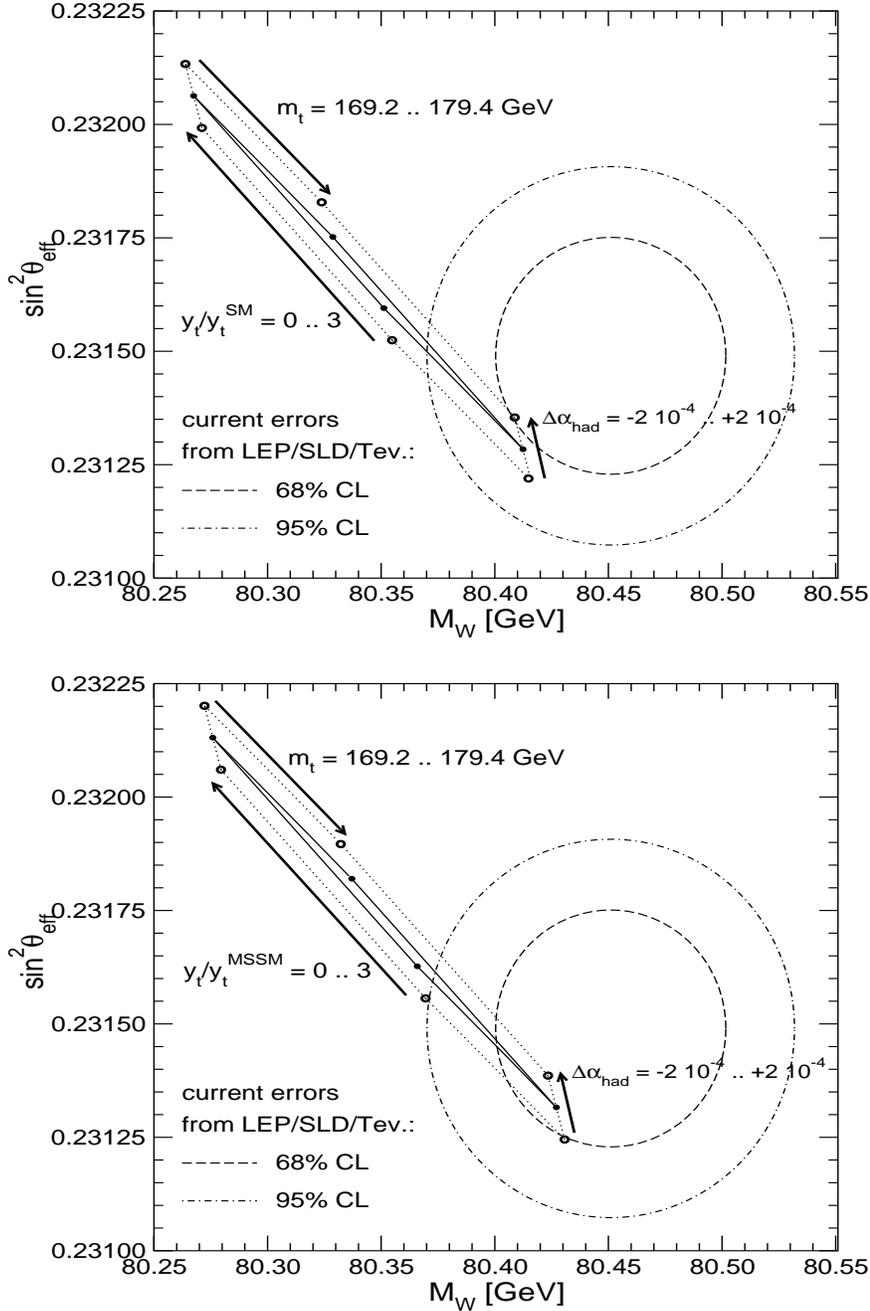

\vspace{-1em}
\begin{center}
\epsfig{figure=SWMWyt02b.bw.eps,width=11.5cm,height=8.5cm} \\[1em]
\epsfig{figure=SWMWyt02bMSSM.bw.eps,width=11.5cm,height=8.5cm} 
\end{center}
\caption[]{
The effect of scaling the top Yukawa 
coupling in the SM (upper plot) and the MSSM (lower plot) for the
precision observables $\MW$ and $\sweff$ is shown in comparison with
the current experimental precision.  
The variation with $\mt$ and $\De\al_{\rm had}$ is shown within their
current experimental errors. 
For the SM evaluation, $\MHSM$ has
been set to the conservative value of $\MHSM = 114 \gev$ (see text). 
For the MSSM evaluation the parameters are $\msusy = 1000 \gev$, 
$\Xt = 2000 \gev$, $\MA = 175 \gev$, $\tb = 3$ and $\mu = 200 \gev$,
resulting in $\mh \approx 114 \gev$. The obtained upper bounds are
$y_t < 1.3 \, y_t^{\SM}$ and $y_t < 1.7 \, y_t^{\MS}$ 
for $\mt = 174.3 \gev$ 
as well as $y_t < 2.2 \, y_t^{\SM}$ and $y_t < 2.5 \, y_t^{\MS}$ 
for $\mt = 179.4 \gev$, all at the 95\% C.L.
}
\label{fig:canyoutellme}
\vspace{-1em}
\end{figure}

\reffi{fig:canyoutellme} shows the effect of varying the top Yukawa 
coupling in the SM and the MSSM for the precision observables $\MW$
and $\sweff$ in comparison with the current experimental precision. 
The allowed 68\% and 95\% C.L.\ contours are indicated in the figure.
The Yukawa coupling is scaled in the following way,
\BE
y_t = x \, y_t^{\SM} , \quad 0 \leq x \leq 3 ,
\EE
and analogously in the MSSM. A shift of this kind in the relation
between a fermion mass and the corresponding Yukawa coupling can occur
for instance in the MSSM (see e.g.\ \citere{MarHowRep}),
\BE
y_t = \frac{\wz \, \mt}{v\, \Sb} \, \ed{1 + \De_t}~,
\EE
where $\De_t$ is induced by SUSY loop corrections. 
Here we do not assume any particular scenario but use
the variation of the top Yukawa coupling only for demonstrating the
sensitivity to this parameter.

For the evaluation of $\MW$ and $\sweff$ in the SM and the MSSM, we
take into account the complete \onel\ results as well as the leading
\twol\ \order{\al\als} and \order{\al_t^2} corrections (as discussed in
\citeres{EWPOSM,EWPOMSSM}). 
Since the SM prediction deviates more from the experimental central
value for increasing values of $\MHSM$, we have chosen in the figure
$\MHSM = 114 \gev$~\cite{LEPHiggsSM} as a conservative value.
The current $1\si$ uncertainties in $\mt$ and $\De\al_{\rm had}$ are
also taken into account, as indicated in the figures. 
Varying the SM top Yukawa coupling (upper plot) yields an upper
bound of $y_t < 1.3 \, y_t^{\SM}$ for $\mt = 174.3 \gev$ 
and of $y_t < 2.2 \, y_t^{\SM}$ for $\mt = 179.4 \gev$, both at the
95\% C.L. These relatively strong bounds are of course related to the
fact that the theory prediction in the SM shows some deviation from
the current experimental central value. 

The lower plot of \reffi{fig:canyoutellme} shows the analogous
analysis in the MSSM for one particular example of SUSY parameters.
We have chosen a large value of $\msusy$, $\msusy = 1000 \gev$,
in order to justify the approximation of neglecting the
\order{\al_t^2} contributions from SUSY loops. The other parameters are
$\Xt = 2000 \gev$, $\MA = 175 \gev$, $\tb = 3$ and $\mu = 200 \gev$,
resulting in $\mh \approx 114 \gev$ (for comparison with the SM
case). The SUSY contributions to $\MW$ and $\sweff$ lead to a somewhat
better agreement between the theory prediction and experiment and
consequently to somewhat weaker bounds on $y_t$. In this example we
find $y_t < 1.7 \, y_t^{\MS}$ for $\mt = 174.3 \gev$ and 
$y_t < 2.5 \, y_t^{\MS}$ for $\mt = 179.4 \gev$, both at the 95\%~C.L. 

In order to demonstrate the sensitivity of future colliders for the
determination of the top Yukawa coupling from electroweak precision
observables, we list in \refta{tab:ytfut} the bounds on $y_t$
obtainable at the LHC and a future LC with GigaZ
option~\cite{gigaz}. Here we assume that the future experimental
central values of $\MW$ and $\sweff$ agree with the theory predictions
for $y_t/y_t^{\SM} = 1$ and $y_t/y_t^{\MS} = 1$, respectively. 
An accuracy in the indirect determination of $y_t$ of about 40\% can
be achieved with the GigaZ precision at the 95\%~C.L.
This is similar to the accuracy achievable from the $t\bar
t$~threshold measurements, see \citere{topthresholdyt}. The results in
\refta{tab:ytfut} are the same for the SM and our SUSY example, since the
only difference (after assuming that the future experimental central
values of $\MW$ and $\sweff$ agree perfectly with the SM or MSSM
predictions) are the relatively small deviations at \order{\al_t^2}
between the SM and the MSSM shown in \reffis{fig:delMW},
\ref{fig:delsw2eff}.

\begin{table}[htb]
\renewcommand{\arraystretch}{1.5}
\BC
\begin{tabular}{|c||c|c|c|}
\cline{2-4} \multicolumn{1}{c||}{}
& LHC ($\de\mt = 2 \gev$) & LHC ($\de\mt = 1 \gev$) & LC/GigaZ \\
\hline\hline
$y_t/y_t^{\SM}$ & 2.5 & 2.3 & 1.4 \\ \hline
$y_t/y_t^{\MS}$ & 2.5 & 2.3 & 1.4 \\ \hline
\end{tabular}
\EC
\renewcommand{\arraystretch}{1}
\caption{
Sensitivity to the top Yukawa coupling at future colliders:
For the LHC we assume $\de\mt = 2 \gev$ or $1 \gev$, $\de\MW = 15 \mev$, 
$\de\sweff = 17 \times 10^{-5}$~\cite{EWPOSM,LHCprec}, while for
LC/GigaZ we use $\de\mt = 0.1 \gev$, $\de\MW = 7 \mev$ and 
$\de\sweff = 1.3 \times 10^{-5}$~\cite{moenig,gigaz,EWPOSM}. 
For $\De\al_{\rm had}$ we
assume a future uncertainty of $\de\De\al_{\rm had} = 5 \times 10^{-5}$.
The bounds on $y_t/y_t^{\SM}$ and $y_t/y_t^{\MS}$ are given at the 
95\%~C.L., assuming that the theory predictions agree with the
experimental central values 
for $\MW$ and $\sweff$ for $y_t/y_t^{\SM,\SU} = 1$. 
The SUSY parameters are chosen as for \reffi{fig:canyoutellme}.
}
\label{tab:ytfut}
\end{table}


\section{Conclusions}

We have calculated the leading \order{\al_t^2}, \order{\al_t \al_b},
and \order{\al_b^2} corrections to  
$\De\rho$ in the MSSM in the limit of heavy squarks. The analytical
results for arbitrary values of the lightest $\cp$-even Higgs boson
mass have been presented. While for the \order{\al_t^2} result all
parameters in the $\cp$-even Higgs sector could be kept arbitrary, for
the full result further tree-level relations had to be employed, which
lead to SM-like couplings of the lightest $\cp$-even Higgs boson to
gauge bosons and SM fermions.  

Numerically we compared the effect of the new MSSM
contribution with  
the leading \order{\al_t^2} SM contribution.
For small $\tb$, it is sufficient to restrict to the \order{\al_t^2}
corrections. Their numerical effect 
is always larger than the \order{\al_t^2} SM contribution.
The corrections to the precision
observables $\MW$ and $\sweff$ amount up to $-12 \mev$ for $\MW$ and about
$+6 \times 10^{-5}$ for $\sweff$. The effective change from
the SM \order{\al_t^2} result with $\MHSM = \mh$ is 
smaller. It amounts up to $-3 \mev$ for $\MW$ and $+2 \times 10^{-5}$
for $\sweff$. This effective change goes to zero for large $\MA$,
i.e. the non-SM contribution decouples.

The \order{\al_t \al_b} and \order{\al_b^2} contributions become
important for large $\tb$. They enter with a different sign than the
\order{\al_t^2} corrections and can overcompensate the latter. 
For large $\tb$ the 
effective change in the predictions for the precision observables
from the whole \order{\al_t^2}, \order{\al_t \al_b}, and
\order{\al_b^2} corrections can amount up to $\de\MW \approx +5 \mev$ 
and $\de\sweff \approx -3\times 10^{-5}$. 
For large $\MA$ also in this case the SM result is recovered.

The MSSM corrections to the electroweak precision observables
discussed here are important in order to reduce the theoretical
uncertainties from unknown higher order corrections within the MSSM to
a similar level as currently reached for the SM.
Achieving this level of theoretical accuracy will be mandatory in
particular in view of the prospective accuracies at a future linear
collider running on the $Z$~peak and the $WW$~threshold.

We have furthermore discussed the sensitivity of the electroweak
precision observables to the top Yukawa coupling, which enters at the
\twol\ level. Varying the SM top Yukawa coupling and requiring
consistency with the present experimental values of $\MW$ and $\sweff$
at the 95\% C.L.\ yields
an upper bound of $y_t < 1.3 \, y_t^{\SM}$ for $\mt = 174.3 \gev$.
This bound can be relaxed within the MSSM, where additional
contributions from SUSY loops to the electroweak precision observables
can lead to a better agreement with the experimental data. 
We have also analyzed the sensitivity of future colliders for the
determination of the top Yukawa coupling from electroweak precision
observables, assuming that the future experimental
central values of $\MW$ and $\sweff$ agree with the theory predictions
for unmodified Yukawa couplings. 
An accuracy in the indirect determination of $y_t$ of about 40\% can
be achieved with GigaZ precision at the 95\%~C.L., which 
is similar to the accuracy achievable from $t\bar t$~threshold
measurements.


\subsection*{Acknowledgements}
\vspace{-.5em}
S.H.\ acknowledges hospitality and financial support by CERN and
IPPP~Durham, where part of the work has been done. 
G.W.\ thanks the Max-Planck Institut f\"ur Physik in Munich for
the hospitality offered to him during the final stage of this work. 
We thank W.~Hollik for useful discussions and T.~Hahn for technical help.
This work has been supported by the European Community's Human
Potential Programme under contract HPRN-CT-2000-00149 Physics at
Colliders.


\begin{appendix}

\section*{Appendix}

\section{Analytical results or arbitrary values of $\mh$}

\subsection{The result for the \order{\al_t^2} contributions for
arbitrary parameters in the $\cp$-even Higgs sector}

We give here the analytical result for the \order{\al_t^2} contributions
where $\mh$, $\mH$ and $\al$ are kept as arbitrary parameters, 
see \refse{subsec:gf2mt4eval}. The full result for the \order{\al_t^2} 
contributions can conveniently be expressed in terms of 
\BE
A \equiv \frac{\mt^2}{\MA^2}, \quad H \equiv \frac{\mt^2}{\mH^2},
 \quad h \equiv \frac{\mt^2}{\mh^2}~.
\EE
\newcommand{\SQA}{\sqrt{1 - 4\,A}\,}
\newcommand{\SQH}{\sqrt{1 - 4\,H}\,}
\newcommand{\SQh}{\sqrt{1 - 4\,h}\,}
We furthermore use the abbreviations 
$s_x \equiv \sin x, c_x \equiv \cos x$ and 
\BEA
\La_x &\equiv& 3 A^2 (1 - 6 x + 10 x^2) - 4 A x (1 - 5 x + 7 x^2)
               + x^2 (1 - 4 x + 6 x^2) \non \\
\La'_x &\equiv& A^2 (-1 + 12 x + 12 x^2) - 2 A x (-1 + 10 x + 4 x^2)
                + x^2 (-1 + 8 x) \non \\
Y_x &\equiv& 2 x (-1 + 4 x) \KKL A (1 - 4 x) - x \KKR \non \\
P_3 &\equiv& 2 A^2 (-1 + 4 A) (h - H) 
           \Big[ 2 A^3 + A^2 (-1 + 2 h + 2 H) - 6 A h H + h H \Big] \non \\
P_4 &\equiv& 2 A (-1 + 4 A) (h - H) 
      \Big[ 20 A^4 - 2 A^3 (5 + 6 h + 6 H) \non \\
 &&       + A^2 (1 + 8 H + 8 h + 4 h H) - A (h + H + 6 h H) + h H \Big] \non \\
P'_4 &\equiv& -2 A (-1 + 4 A) (h - H) 
      \Big[18 A^4 - A^3 (9 + 14 h + 14 H) \non \\
 &&      + A^2 (1 + 8 H + 8 h + 10 h H)
         - A (h + H + 7 h H) + h H \Big] \non \\
P_5 &\equiv& 4 A^5 + A^4 (54 + 136 h) - 2 A^3 (17 + 108 h + 54 h^2) \non \\
 &&                 + A^2 (5 + 94 h +146 h^2) - 2 A h (6 + 29 h) + 7 h^2 ~.
\EEA

\noindent
The \twol\ contribution to the \rp\ then reads:
\BEA
\lefteqn{
\De\rho_{1,\hi}^{\SU} = 3 \, \frac{\gf^2}{128 \,\pi^4} \, \mt^4  
                      \ed{\sbe^2}\;\times } \\
 && \Bigg\{
    \Liz{(1-\SQH)/2} \frac{2}{\SQH (A - H)^2 H^2} 
     \Bigg[ \sa^2 \La_H + \sbe \sa (\sbe \sa + \cbe \ca) Y_H \non \\
 &&    \KKL \sbe \cbe (\sbe \cbe + \sa \ca) 
       + \sbe \sa (\sbe \sa + \cbe \ca) 
       + 2 \sbe^3 \sa (\sbe \sa + \cbe \ca)\KKR 2 H^3 (-1 + 4 H)
       +  \Bigg] \non  \\  
 &+&\Liz{(1 - \SQh)/2} \frac{2}{\SQh (A - h)^2 h^2}
     \Bigg[ \ca^2 \La_h + \sbe \ca (\sbe \ca - \sa \cbe) Y_h \non \\
 &&    - \KKL - 2 \sa \ca \sbe \cbe (1 + \sbe^2) + \sbe^2 (1 + \ca^2)
    + \sbe^4 (1 - 2 \sa^2)\KKR  2 h^3 (1 - 4 h) \Bigg] \non \\
 &+&\Liz{(1 - \SQA)/2} 
     \frac{2 \cbe}{\SQA (1 - 4 A) A^2 (A - h)^2 (A - H)^2} \non \\
 &&  \Bigg[ \sa \ca \sbe P_4 
    - \KKL 2 \sa \ca \sbe^3 + \sbe^2 \cbe (1 - 2 \sa^2)\KKR P_3
   + \cbe (A - H)^2 P_5
   + \cbe \sa^2 P'_4 \Bigg] \non \\
 &+&\Liz{1-\ed{H}} \frac{(H - 1)^2}{(A - H)^2 H^2}
     \Bigg[\sa^2 \KKL A^2 (-3 + 6 H) - 4 A (-1 + H) H - H^2 \KKR \non \\
 &&  + \KKL \sbe^2 (1 + \sa^2) - \sbe^4 (1 - 2 \sa^2) 
     + 2 \sbe \cbe \sa \ca (1 + \sbe^2) \KKR 2 H^3 \non \\
 &&  + \sbe \sa (\sbe \sa + \cbe \ca) 2 H (A (1 - 4 H) - H) \Bigg] \non \\ 
 &-&\Liz{1-\ed{h}} \frac{(h - 1)^2}{(A - h)^2 h^2}
     \Bigg[ \KKL 2 \sbe \cbe (1 + \sbe^2) \sa \ca - \sbe^2  (1 + \ca^2) 
           - \sbe^4  (1 - 2 \sa^2) \KKR 2 h^3 \non \\
 &&        + \sbe \ca (\cbe \sa - \sbe \ca) 2 h (A (1 - 4 h) - h)
           + \ca^2 \KKL A^2 (3 - 6 h) + 4 A h (-1 + h) + h^2\KKR \Bigg] \non \\
 &-&\Liz{1-\ed{A}} \frac{\cbe}{A^2 (A - h)^2 (A - H)^2}
     \Bigg[ \sa \ca \sbe 2 (A - 1) A (h - H) 
     \Big[ 2 A^4 - 7 A^3  \non \\
 && + A^2 (1 + 5 h + 5 H - 2 h H) 
      - A (h + H + 3 h H) + h H\Big]  \non \\
 && + \KKL 2 \sa \ca \sbe^3 + \sbe^2 \cbe (1 - 2 \sa^2)\KKR  
     2 (A - 1) A^2 (h - H) 
     \Big[ A^3 + A^2 (1 - 2 h - 2 H)  \non \\
 && + 3 A h H - h H\Big]
  + \cbe (A - H)^2 
     \Big[ A^4 (-3 + 6 h) - 4 A^3 (1 + 3 h + h^2)   \non \\
 && + A^2 (3 + 22h + 11 h^2) - 8 A h (1 + 2 h) + 5 h^2\Big]
  + \cbe \sa^2 (-2) (A - 1) A (h - H) \non \\
 &&   \Big[ 3 A^4 - 2 A^3 (3 + h + H) + A^2 (1 + 5 h + 5 H + h H)
      - A (h + H + 4 h H) + h H\Big] \Bigg] \non \\
 &+&\log^2(h) \ed{2 \SQh (A - h)^2 h^2}
     \Bigg[ \Big\{ 2 \sbe \sa \KKL \sbe \sa + \cbe \ca (1 + \sbe^2)\KKR \non \\
 &&  - \sbe^2 \KKL 2 + \sa^2 + \sbe^2 (1 - 2 \sa^2)\KKR \Big\} 2 h^3 (1 - 4 h) 
    + \sbe \sa \ca \cbe 2 h (-1 + 4 h) (A (-1 + 4 h) + h) \non \\
 && + \ca^2 \La_h
    + \sbe^2 \ca^2 Y_h \Bigg] \non \\
 &+&\log^2(H) \ed{2 \SQH (A - H)^2 H^2}
    \Bigg[ \sa^2 \La_H
  + \Big\{ \sbe^2 \KKL 1 + \sa^2 - \sbe^2 (1 - 2 \sa^2)\KKR + \non \\
 &&  2 \sbe \cbe \sa \ca (1 + \sbe^2)\Big\} 2 H^3 (-1 + 4 H) 
  + \sbe \sa (\sbe \sa + \cbe \ca) Y_H \Bigg] \non \\
 &+&\log^2(A) \frac{\cbe}{2 (1 - 4 A)^2 A^2 (A - h)^2 (A - H)^2}
     \Bigg[ \sbe \sa \ca \SQA P_4 \non \\
 && + \sbe^2 \KKL -2 \sbe \sa \ca - \cbe (1 - 2 \sa^2)\KKR \SQA P_3
    + \cbe (A - H)^2 \Big[ 2 (1 - 4 A)^2 (A - h)^2   \non \\
 && + \SQA P_5 \Big]
    + \cbe \sa^2 \SQA P'_4 \Bigg] \non \\
 &+&\log(h) \ed{2 h^2 (A - h)^2}
     \Bigg[ \KKL - 2 \sbe \cbe (1 + \sbe^2) \sa \ca + \sbe^2 \ca^2
    + \sbe^2 (1 + \sbe^2 (1 - 2 \sa^2)) \KKR 4 h^4 \non \\
 && + \sbe \ca (\cbe \sa - \sbe \ca) 4 h^2 (A (1 + 4 h) - h)
    + \ca^2 \La'_h \Bigg] \non \\
 &+&\log(H) \ed{2 H^2 (A - H)^2}
     \Bigg[ \sa^2 \La'_H
    + \Big[ \sbe^2 + \sbe \sa (\sbe \sa + \cbe \ca) - \sbe^4 (1 - 2 \sa^2)
       \non \\
 && + \sbe \cbe \sa \ca (1 + 2 \sbe^2)\Big] 4 H^4
    + \sbe \sa (\sbe \sa + \cbe \ca) 4 H^2 (-A (1 + 4 H) + H) \Bigg] \non \\
 &+&\log(A) \ed{2 A^2 (-1 + 4 A) (A - h)^2 (A - H)^2}
     \Bigg[ -(A - H)^2   \Big[ A^4 (-76 + 48 h)  \non \\
 && + A^3 (32 + 156 h - 32 h^2)
       + A^2 (-3 - 68h - 84 h^2) + 6 A h (1 + 6 h) - 3 h^2\Big] \non \\
 && + \sa^2 4 A^2 (-1 + 4 A) (h - H)  
      \KKL 3 A^3 + A^2 (1 - 2 h - 2 H) - A (h + H - h H) + h H \KKR \non \\
 && + \sbe^2 \Big[ 4 A^6 (-19 + 8 h + 4 H) 
              - 4 A^5 (-8 -40 h - 37 H + 24 h H + 8 H^2)   \non \\
 &&          + A^4 (-3 - 64 H - 68 h - 68 H^2 - 92 h^2 - 312 h H
                 + 16 h^2 H + 96 h H^2)  \non \\
 &&          + A^3 (6 H + 6 h + 32 H^2 + 36 h^2 + 136 h H 
                 + 180 h^2 H + 144 h H^2 - 32 h^2 H^2)  \non \\
 &&          + A^2 (-3 h^2 - 3 H^2 - 12 h H - 72 h^2 H - 68 h H^2 - 84 h^2 H^2)
          + 6 A h H (h + H + 6 h H)  \non \\
 && - 3 h^2 H^2\Big]
  + \sbe^2 \sa^2 (-4 A^2) (-1 + 4 A) (h - H)  
      \Big[ A^3 + A^2 (1 + 2 h + 2 H)  \non \\
 && - A (h + H + 5 h H) + h H \Big]
  + \sbe^4 (1 - 2 \sa^2) 4 A^3 (-1 + 4 A) (h - H) 
    \Big[ A^2 - 2 A (h + H)  \non \\
 && + 3 h H\Big]
  + \sbe \cbe \sa \ca (-4 A^2) (-1 + 4 A) (h - H)  
      \KKL 2 A^3 + A^2 - A(h + H + 2 h H) + h H \KKR  \non \\
 &&  + \sbe^3 \cbe \sa \ca (-8 A^3) (-1 + 4 A) (h - H) 
       \KKL A^2 - 2 A (h + H) + 3 h H\KKR
  \Bigg] \non \\
 &+&\log\KKL(1-\SQh)/2\KKR \ed{\SQh (A - h)^2 h^2}
     \Bigg[ - \ca^2 \La_h
  + \Big[-2 \sbe^2 \non \\
 &&  + \sbe \sa (\sbe \sa + \cbe \ca)  - \sbe^4 (1 - 2 \sa^2) 
     + \sbe \cbe \sa \ca (1 + 2 \sbe^2)\Big] 2 h^3 (-1 + 4 h) \non \\
 && + \KKL \sbe \sa (\sbe \sa + \cbe \ca) - \sbe^2\KKR Y_h \Bigg] \non \\
 &-&\log\KKL(1-\SQH)/2\KKR \ed{\SQH (A - H)^2 H^2}
     \Bigg[ \sa^2 \La_H
  + \Big[ \sbe^2 (1 + \sa^2)  \non \\
 && - \sbe^4 (1 - 2 \sa^2) 
    + 2 \sbe \cbe \sa \ca (1 + \sbe^2)\Big] 2 H^3 (-1 + 4 H)
  + \sbe \sa (\sbe \sa + \cbe \ca) Y_H \Bigg] \non \\
 &-& \log\KKL(1-\SQA)/2\KKR 
      \frac{\cbe}{(1 - 4 A) \SQA A^2 (A - h)^2 (A - H)^2}
      \Bigg[  \sbe \sa \ca P_4 \non \\
 &&  + \sbe^2 \KKL -2 \sbe \sa \ca - \cbe (1 - 2 \sa^2)\KKR P_3
  + \cbe (A - H)^2 P_5
  + \cbe \sa^2 P'_4 \Bigg] \non \\  
 &+& \log\KL\frac{1-\SQh}{1+\SQh}\KR 
      \frac{\ca^2 \SQh (1-4 h)}{2 h^2} \non \\
 &+& \log\KL\frac{1-\SQH}{1+\SQH}\KR 
      \frac{\sa^2 \SQH (1-4 H)}{2 H^2} \non \\
 &+& \log\KL\frac{1-\SQA}{1+\SQA}\KR 
      \frac{\cbe^2 \SQA}{2 A^2} \non \\
 &+& \log\KL|-1 + 1/A|\KR \frac{-\cbe^2 (A - 1)^2}{A^2} \non \\
 &+& \ed{6 (1 - 4 A)^2 A^2 h^2 H^2 (A - h)^2 (A - H)^2 
               (-1 + 4 h) (-1 + 4 H)} \times \non \\
 && \Bigg[ 6 (1 - 4 A)^2 A h H (A - h) (A - H) (-1 + 4 h) (-1 + 4 H) 
    \times   \non \\
 &&     \Bigg( H (A - H) \KKL A^2 (-4 + 25 h) - A h (7 + 23 h) + 11 h^2\KKR
     + \sa^2 (-2 A) (h - H)  \non \\
 && \KKL 2 A^2 + A (-2 H - 2 h + h H) + 2 h H \KKR
     + \sbe^2 h H \Big[ A^2 (11 - 4 h + 2 H)  \non \\
 && + A (-11 H - 11 h + 2 h H) + 11 h H\Big]
     + \sbe^2 \sa^2 H h (6 A^2) (h - H) \non \\
 &&  + \sbe^4 (1 - 2 \sa^2) (2 A^2 h H (h - H))
     + \sbe \cbe^3 \sa \ca 4 A^2 h H (h - H) \Bigg) \non \\
 &+& \pi^2 \Bigg(
     (1 - 4 A)^2 (1 - 4 h) (1 - 4 H) (A - h)^2 (A - H)^2 h^2 H^2  
       \Big\{ 2 A^2 (h - H) \Big[ (-1 + 2 \sa^2) \sbe^4  \non \\
 &&     + 2 \sa \ca \sbe (1 + \sbe^2) \cbe\Big]
        + 2 + \sbe^2 \KKL -2 + A^2 \{ 3 + 2 h (-2 + \sa^2) - 2 H (1 +
               \sa^2)\} \KKR \Big\} \non \\
 &&    + (1 - 4 A)^2 A^2 h^2 (A - h)^2 (-1 + 4 h) \SQH   
       \Big\{  \sa^2 \La_H
        + \sbe \sa (\sbe \sa + \cbe \ca) Y_H \non \\
 &&     + \KKL \sbe^2 (1 + \sa^2) - \sbe^4 (1 - 2 \sa^2) 
           + 2 \sbe \cbe \sa \ca (1 + \sbe^2)\KKR 2 H^3 (-1 + 4 H) 
       \Big\} \non \\
 &&    - (1 - 4 A)^2 A^2 H^2 (A - H)^2 (-1 + 4 H) \SQh   
       \Big\{- \ca^2 \La_h
        + \sbe \ca (-\sbe \ca + \cbe \sa) Y_h \non \\
 &&     + \KKL \sbe^2 \sa^2 - 2 \sbe^2 - \sbe^4 (1 - 2 \sa^2) 
           + 2 \sbe \cbe \sa \ca (1 + \sbe^2)\KKR 2 h^3 (-1 + 4 h) 
       \Big\} \non \\
 &&    - \KKL \SQA h^2 H^2 (1 - 4 h) (1 - 4 H)\KKR
       \Big\{  (A - H)^2 P_5
        + \sa^2 P'_4
        + \sbe^2 \Big[ -4 A^7  \non \\
 && + A^6 (-54 - 152 h + 24 H) 
                + 2 A^5 (17 + 114 h + 48 H + 46 h^2 + 6 H^2 + 136 h H)
               \non \\
 &&               - A^4 (5 + 96 h + 66 H + 142 h^2 + 58 H^2 + 432 h H
                       + 168 h^2 H + 184 h H^2) \non \\
 &&             + 2 A^3 (6 h + 5 H + 29 h^2 + 17 H^2 + 94 h H
                         + 136 h^2 H + 118 h H^2 + 54 h^2 H^2) \non \\
 &&             - A^2 (7 h^2 + 5 H^2 + 24 h H + 114 h^2 H + 96 h H^2
                       + 146 h^2 H^2) \non \\
 &&             + 2 A h H (6 H + 7 h + 29 h H) - 7 h^2 H^2 \Big]
        + \sbe^2 \sa^2 2 A (-1 + 4 A) (h - H)  
            \Big[ 22 A^4  \non \\
 && - A^3 (11 + 10 h + 10 H) + A^2 (1 + 8 h + 8 H - 2 h H)
             - A (h + H + 5 h H) + h H\Big] \non \\
 &&     + \sbe^3 (\sbe - 2 \sbe \sa^2 - 2 \cbe \sa \ca) P_3
        + \sbe \cbe \sa \ca P_4 \Big\} \Bigg)
\Bigg]
\Bigg\} . \non
\label{drfull}
\EEA

\subsection{The result for the \order{\al_t^2}, \order{\al_t\al_b} and
\order{\al_b^2} corrections}

In the following we list the full result for the \order{\al_t^2},
\order{\al_t\al_b} and \order{\al_b^2} corrections. As explained in 
\refse{subsec:gf2mt4eval}, it has been obtained by using the Higgs sector 
relations \refeqs{chargedHiggsMW0} and (\ref{heavyHiggsMW0}). We give
this result for arbitrary space--time dimension $D$, using the shorthands
\BE
\De_{i, j, k, l} = i + j \, D + k \, D^2 + l \, D^3, \quad
\De_{i, j, k} = i + j \, D + k \, D^2~.
\EE
The result is expressed in terms
of the one-loop scalar integrals $A_0(m)$ and $B_0(p^2,m_1,m_2)$ as
defined in \citere{dr2lA} and the scalar two-loop vacuum integral 
$T_{134}(m_1^2,m_2^2,m_3^2)$ as defined in \citere{t134}.

\def\re{\mathop{\mathrm{Re}}}
\begin{align}
& \De\rho_{1, {\rm Higgs}}^{\SU, \al_{t,b}^2} \; = \;
\frac{3 \, \gf^2}{(4 \pi)^4 D} \times
\\&{}\notag
\Bigg[ \; A_0(\mt)\,
   \Bigg\{ 8\,(D - 2) \,\mt^2\,
      B_0(\mb^2,0,\mt)
\\&{}\notag
    - \frac{16\,(D - 2) \,\mb^2\,
        \left( \MA^2 - 2\,\mb^2 \right) \,\mt^2\,
        \sbe^2\,B_0(\mb^2,\MA,\mb)}
{\cbe^2\,{\left( \mb^2 - \mt^2 \right) }^2} 
    + \frac{8\,(D - 2) \,\mt^2}{\cbe^2\,
        {\left( \mb^2 - \mt^2 \right) }^2\,\sbe^2}  
\\&{}\notag
        \Big( \cbe^4\,\mt^4 + \mb^4\,\sbe^4 + 
          \mb^2\,\mt^2\,
           \left( 1 + 2\,\sbe^2 - 2\,\sbe^4 \right)  - 
          \MA^2\,\left( \cbe^4\,\mt^2 + 
             \mb^2\,\sbe^4 \right)  \Big) \,
        B_0(\mb^2,\MA,\mt)
\\&{}\notag
    + \frac{8\,(D - 2) \,\mb^2\,
        \left( 4\,\mb^2 - m_{h}^2 \right) \,\mt^2\,
        B_0(\mb^2,\mb,m_{h})}{{\left( m
            _{b}^2 - \mt^2 \right) }^2}
\\&{}\notag
    - \frac{(D - 2) \,
        \re(B_0(\mt^2,0,\mb))\,
        \left( (D - 2) \,D\,\mb^4 - 
          2\,(D - 4) \,D\,\mb^2\,\mt^2 + 
          \mt^4\,\De_{8, -6, 1} \right) }{\mt^2}
\\&{}\notag
    - \frac{(D - 2)}{\cbe^2\,
        {\left( \mb^2\,\mt - \mt^3 \right) }^2\,\sbe^2} 
       \Big[ \Big( \cbe^4\,\mt^4 + \mb^4\,\sbe^4 + 
          \mb^2\,\mt^2\,
           \left( 1 + 2\,\sbe^2 - 2\,\sbe^4 \right)  - 
          \MA^2\,\left( \cbe^4\,\mt^2 + 
             \mb^2\,\sbe^4 \right)  \Big) \,
\\&{}\notag
        \re(B_0(\mt^2,\MA,\mb))\,
        \left( (D - 2) \,D\,\mb^4 - 
          2\,(D - 4) \,D\,\mb^2\,\mt^2 + 
          \mt^4\,\De_{8, -6, 1} \right) \Big]
\\&{}\notag
     + \frac{2\,\cbe^2\,(D - 2) \,
        \left( \MA^2 - 2\,\mt^2 \right) \,
        B_0(\mt^2,\MA,\mt)}
            {{\left( \mb^2 - 
            \mt^2 \right) }^2\,\sbe^2} 
        \Big( (D - 2) \,D\,\mb^4 - 
          2\,(D - 4) \,D\,\mb^2\,\mt^2  
\\&{}\notag
        +  \mt^4\,\De_{8, -6, 1} \Big) 
      + \frac{(D - 2) \,
        \left( m_{h}^2 - 4\,\mt^2 \right) \,
        B_0(\mt^2,m_{h},\mt)}
             {{\left( \mb^2 - 
               \mt^2 \right) }^2} \Bigg\} 
\\&{}\notag
        \left( (D - 2) \,D\,\mb^4 - 
          2\,(D - 4) \,D\,\mb^2\,\mt^2 + 
          \mt^4\,\De_{8, -6, 1} \right)
\end{align}

\removelastskip

\begin{align}
\notag
&  + A_0(\mh)\,A_0(\mt)\,
\frac{1}{\mh^2\,
     \left( \mh - 2\,\mt \right) \,\left( \mh + 2\,\mt \right) \,
     \left( -\mb^2 + \mt^2 \right) } 
     \Big( -48\,\mb^4\,\left( \mh^2 - 4\,\mt^2 \right)   
\\&{}\notag
      + \mt^2\,\left( 192\,\mt^4 + 
          4\,\mh^2\,\mt^2\,\De_{-40, 24, -13, 2} - 
          \mh^4\,\De_{-24, 16, -8, 1} \right)
\\&{}\notag
       + \mb^2\,\left( -384\,\mt^4 - 
          4\,\mh^2\,\mt^2\,\De_{-20, 16, -9, 2} + 
          \mh^4\,\De_{8, 8, -4, 1} \right)  \Big)
\end{align}

\removelastskip

\begin{align}
\notag
&  + \frac{2\,{A_0(\mt)}^2}{\cbe^2\,\left( -3 + D \right) \,
     \left( \MA - 2\,\mt \right) \,\left( \MA + 2\,\mt \right) \,
     \left( -\mh + 2\,\mt \right) \,\left( \mh + 2\,\mt \right) \,
     \left( -\mb^2 + \mt^2 \right) \,\sbe^2}
\\&{}\notag
       \Bigg\{ \Big[ -
        2\,\mt^2\,\Bigg( (D - 2) \,\mb^2\,
            \Bigg( -4\,\mt^2\,
                \left( \sbe^2\,
                   \left( -3\,D^3 + \De_{-58, 17, 10} \right)  + 
                  \sbe^4\,\De_{22, -17, 3} + 
                  \De_{12, 8, -13, 3} \right)   
\\&{}\notag
            +  \mh^2\,\left( \sbe^2\,
                  \left( -4\,D^3 + \De_{-52, 6, 16} \right)  + 
                 \De_{12, 8, -13, 3} + 
                 \sbe^4\,\De_{16, -6, -3, 1} \right)  \Bigg)  
\\&{}\notag
          - \cbe^2\,\mt^2\,
            \Big( \mh^2\,
               \left( 3\,D^4 + 
                 \sbe^2\,
                  \left( 5\,D^3 - D^4 + \De_{128, -108, 16} \right)  + 
                 \De_{-72, 60, 30, -23} \right)  
\\&{}\notag
              + 4\,\mt^2\,\left( -3\,D^4 + 
                 \sbe^2\,
                  \left( 3\,D^3 + \De_{-140, 136, -39} \right)  + 
                 \De_{72, -60, -30, 23} \right)  \Big)  \Bigg)   
\\&{}\notag
       + \MA^2\,\Bigg( \mh^2\,
           \Big( -\cbe^2\,\mt^2\,
               \left( D^4 + \sbe^2\,
                  \left( 2\,D^3 + \De_{52, -32, -1} \right)  + 
                 \De_{-24, 8, 24, -11} \right)  
\\&{}\notag
            + (D - 2) \,\mb^2\,
              \left( \left( -10 + 3\,D \right) \,\sbe^4 + 
                D\,\De_{6, -5, 1} - \sbe^2\,\De_{2, 5, -5, 1} 
                                              \right)  \Big)  
\\&{}\notag
           + 2\,\mt^2 \Big( (D - 2) \,\mb^2\,
              \left( -2\,D\,\De_{6, -5, 1} + 
                \sbe^2\,\De_{10, -1, -4, 1} + 
                \sbe^4\,\De_{14, 5, -6, 1} \right) 
\\&{}\notag
            - \mt^2\,\left( \sbe^2\,
                 \left( D^4 + \De_{-164, 108, 27, -18} \right)  - 
                2\,\left( D^4 + \De_{-24, 8, 24, -11} \right)  + 
                \sbe^4\,
                 \left( D^4 + \De_{116, -92, 21, -4} \right)  \right)  
\Big)  \Bigg)  \Big] \Bigg\}
\end{align}

\removelastskip

\begin{align}
\notag
&  + \frac{2\,{A_0(\mb)}^2}{\cbe^2\,\left( -3 + D \right) \,
     \left( \MA - 2\,\mb \right) \,\left( \MA + 2\,\mb \right) \,
     \left( 2\,\mb - \mh \right) \,\left( 2\,\mb + \mh \right) \,
     \left( \mb^2 - \mt^2 \right) \,\sbe^2} 
\\&{}\notag
    \Bigg\{ \MA^2\,
        \Big[ 2\,\mb^4\,\sbe^2\,
           \left( -3\,D^4 + \sbe^2\,
              \left( -4\,D^3 + D^4 + \De_{116, -92, 21} \right)  + 
             \De_{-68, 76, -69, 26} \right)  
\\&{}\notag
          - (D - 2) \,\mh^2\,\mt^2\,
           \left( 4\,\left( -3 + D \right)  + 
             \left( -10 + 3\,D \right) \,\sbe^4 + 
             \sbe^2\,\De_{22, -1, -5, 1} \right)   
\\&{}\notag
         + \mb^2\,\Big( \mh^2\,\sbe^2\,
              \left( D^4 + \sbe^2\,
                 \left( -2\,D^3 + \De_{-52, 32, 1} \right)  + 
                \De_{28, -24, 23, -9} \right)   
\\&{}\notag
            - 2\,(D - 2) \,\mt^2\,
              \left( -8\,\left( -3 + D \right)  + 
                \sbe^4\,\De_{14, 5, -6, 1} - 
                \sbe^2\,\De_{38, 9, -16, 3} \right)  \Big)
                \Big]  
\\&{}\notag
         + 2\,\mb^2\,\Big[ (D - 2) \,\mh^2\,
           \mt^2\,\left( 8\,\left( -3 + D \right)  + 
             2\,\sbe^2\,\De_{10, 3, -5, 1} + 
             \sbe^4\,\De_{16, -6, -3, 1} \right)   
\\&{}\notag
         + 4\,\mb^4\,\sbe^2\,
           \left( 3\,D^4 + \sbe^2\,
              \left( 3\,D^3 + \De_{-140, 136, -39} \right)  + 
             \De_{68, -76, 69, -26} \right)   
\\&{}\notag
         + \mb^2\,\Big( -4\,(D - 2) \,\mt^2\,
               \left( 8\,\left( -3 + D \right)  + 
                 \sbe^4\,\De_{22, -17, 3} + 
                 \sbe^2\,\De_{14, 17, -16, 3} \right)   
\\&{}\notag
            + \mh^2\,\sbe^2\,
              \left( -2\,\left( D^4 + \De_{28, -24, 23, -9} \right)  + 
                \sbe^2\,
                 \left( -D^4 + \De_{128, -108, 16, 5} \right)  \right)  
\Big)  \Big]  \Bigg\} 
\end{align}

\removelastskip

\begin{align}
\notag
&  + A_0(\mb)\,\Bigg\{ (D - 2) \,D\,
      \left( \frac{2 - D}{\mb^2} - 
        \frac{4\,\left( \left( -1 + \frac{2}{D} \right) \,\mb^2 + 
             \mt^2 \right) }{{\left( \mb^2 - \mt^2 \right) }^2}
                                                         \right) 
\\&{}\notag
    \left( 4\,\mb^2\,\mt^2 - 
        \frac{\left( \MA^2 - \mb^2 - \mt^2 \right) \,
           \left( \cbe^4\,\mt^2 + \mb^2\,\sbe^4 
\right) }{\cbe^2\,\sbe^2} \right) \,
      B_0(\mb^2,\MA,\mt) 
\\&{}\notag
     + 8\,(D - 2) \,\mb^2\,
      \re(B_0(\mt^2,0,\mb))  
     + \frac{8\,(D - 2) \,\mb^2}{\cbe^2\,
        {\left( \mb^2 - \mt^2 \right) }^2\,\sbe^2} \,
        \Big( \cbe^4\,\mt^4 + \mb^4\,\sbe^4  
\\&{}\notag
         + \mb^2\,\mt^2\,
           \left( 1 + 2\,\sbe^2 - 2\,\sbe^4 \right)  - 
          \MA^2\,\left( \cbe^4\,\mt^2 + 
             \mb^2\,\sbe^4 \right)  \Big) \,
        \re(B_0(\mt^2,\MA,\mb))
\\&{}\notag
    - \frac{16\,\cbe^2\,(D - 2) \,\mb^2\,
        \mt^2\,\left( \MA^2 - 2\,\mt^2 \right) \,
        B_0(\mt^2,\MA,\mt)}{
        {\left( \mb^2 - \mt^2 \right) }^2\,\sbe^2} 
\\&{}\notag
     - \frac{8\,(D - 2) \,\mb^2\,\mt^2\,
        \left( \mh^2 - 4\,\mt^2 \right) \,
        B_0(\mt^2,\mh,\mt)}{{\left( \mb^2 - \mt^2 \right) }^2}  
\\&{}\notag
     - \frac{(D - 2) \,
        B_0(\mb^2,0,\mt)\,
        \left( -2\,(D - 4) \,D\,\mb^2\,\mt^2 + 
          (D - 2) \,D\,\mt^4 + 
          \mb^4\,\De_{8, -6, 1} \right) }{\mb^2} 
\\&{}\notag
     + \frac{2\,(D - 2) \,
        \left( \MA^2 - 2\,\mb^2 \right) \,\sbe^2\,
        B_0(\mb^2,\MA,\mb)}{\cbe^2\,
        {\left( \mb^2 - \mt^2 \right) }^2} 
        \Big( -2\,(D - 4) \,D\,\mb^2\,\mt^2  
\\&{}\notag
          + (D - 2) \,D\,\mt^4 + 
          \mb^4\,\De_{8, -6, 1} \Big) 
     - \frac{(D - 2) \,
        \left( 4\,\mb^2 - \mh^2 \right) \,
        B_0(\mb^2,\mb,\mh)}
        {{\left( \mb^2 - \mt^2 \right) }^2}
\\&{}\notag
        \Big( -2\,(D - 4) \,D\,\mb^2\,\mt^2 + 
          (D - 2) \,D\,\mt^4 + 
          \mb^4\,\De_{8, -6, 1} \Big)
     - \frac{A_0(\mh)}
           {\mh^2\,
        \left( 4\,\mb^2 - \mh^2 \right) \,
        \left( \mb^2 - \mt^2 \right) } 
\\&{}\notag
          \Big( 192\,\mb^6 + 
          4\,\mb^4\,\left( -96\,\mt^2 + 
             \mh^2\,\De_{-40, 24, -13, 2} \right)  + 
          \mb^2\,\big( 192\,\mt^4 - 
             \mh^4\,\De_{-24, 16, -8, 1} 
\\&{}\notag
             - 4\,\mh^2\,\mt^2\,\De_{-20, 16, -9, 2} \big)   
         + \mh^2\,\mt^2\,
           \left( -48\,\mt^2 + \mh^2\,\De_{8, 8, -4, 1} \right) \Big) 
\\&{}\notag
     + \frac{A_0(\mt)}
{\sbe^2\cbe^2\,\mb^2\,\mt^2\,
        {\left( \MA^2 - (\mt - \mb)^2 \right) }^2\,
        {\left( \MA^2 - (\mt + \mb)^2 \right) }^2\,
        \left( \mb^2 - \mt^2 \right)} 
\\&{}\notag
   \Big[ {\left( \mb^2 - \mt^2 \right) }^3\,
           \Bigg( \cbe^2\,{(D - 2) }^2\,D\,
              \mt^8 + {(D - 2) }^2\,D\,\mb^8\,
              \sbe^2 + 
             \mb^4\,\mt^4\,
              \Big( -{\left( 10 - 3\,D \right) }^2\,D 
\\&{}\notag
                + 96\,(D - 2) \,\sbe^2 - 
                96\,(D - 2) \,\sbe^4 \Big)  + 
             \mb^6\,\mt^2\,
              \left( {(D - 2) }^2\,D + 
                48\,(D - 2) \,\sbe^4 - 
                2\,\sbe^2\,\De_{-48, 60, -32, 5} \right)   
\\&{}\notag
            + \mb^2\,\mt^6\,
              \left( 48\,(D - 2) \,\sbe^4 + 
                D\,\De_{-68, 60, -9} + 
                2\,\sbe^2\,\De_{48, 12, -32, 5} \right)  \Bigg)  
\\&{}\notag
         + \MA^8\,\Bigg( -\cbe^2\,{(D - 2) }^2\,D\,
               \mt^6 + {(D - 2) }^2\,D\,\mb^6\,
              \sbe^2 + 
             \mb^2\,\mt^4\,
              \Big( -32\,\left( -1 + D \right)  + 
                \sbe^2\,\De_{-160, 188, -68, 11}  
\\&{}\notag
               - 4\,\sbe^4\,\De_{-32, 40, -18, 3} \Big)  + 
             \mb^4\,\mt^2\,
              \left( {(D - 2) }^2\,D + 
                4\,\sbe^4\,\De_{-32, 40, -18, 3} + 
                \sbe^2\,\De_{96, -132, 76, -13} \right)  \Bigg)  
\\&{}\notag
        - 4\,\MA^6\,\Bigg( -\cbe^2\,{(D - 2) }^2\,D\,
               \mt^8 + {(D - 2) }^2\,D\,\mb^8\,
              \sbe^2 - 
             4\,\mb^4\,\mt^4\,
              \left( -1 + 2\,\sbe^2 \right) \,\De_{8, -10, 1}  
\\&{}\notag
            + \mb^2\,\mt^6\,
              \left( -4\,\De_{-4, 4, 1} + 
                \sbe^2\,\De_{-128, 148, -52, 9} - 
                2\,\sbe^4\,\De_{-56, 68, -30, 5} \right)   
\\&{}\notag
            + \mb^6\,\mt^2\,
              \Big( {(D - 2) }^2\,D + 
                2\,\sbe^4\,\De_{-56, 68, -30, 5}  
               + \sbe^2\,\De_{96, -124, 68, -11} \Big)  \Bigg)  
\\&{}\notag
        - 4\,\MA^2\,{\left( \mb^2 - \mt^2 \right) }^2\,
           \Bigg( -\cbe^2\,{(D - 2) }^2\,D\,
               \mt^8 + {(D - 2) }^2\,D\,\mb^8\,
              \sbe^2 
\\&{}\notag
            + 4\,\mb^4\,\mt^4\,
              \left( -1 + 2\,\sbe^2 \right) \,
              \De_{-4, 18, -8, 1}             
            + \mb^6\,\mt^2\,
              \left( {(D - 2) }^2\,D + 
                4\,\sbe^4\,\De_{-24, 20, -6, 1} + 
                \sbe^2\,\De_{96, -116, 60, -9} \right)   
\\&{}\notag
            + \mb^2\,\mt^6\,
              \left( 4\,D\,\De_{8, -8, 1} - 
                4\,\sbe^4\,\De_{-24, 20, -6, 1} - 
                \sbe^2\,\De_{96, -44, -12, 1} \right)  \Bigg)   
\\&{}\notag
            +           2\,\MA^4\,\Bigg( -
              3\,\cbe^2\,{(D - 2) }^2\,D\,\mt^{10} + 
             3\,{(D - 2) }^2\,D\,\mb^{10}\,\sbe^2 - 
             2\,\mb^6\,\mt^4\,
              \Big( \sbe^4\,
                 \left( 17\,D^3 + \De_{-72, 204, -118} \right)  
\\&{}\notag
                - 4\,\sbe^2\,
                 \left( 5\,D^3 + \De_{-32, 78, -36} \right)  + 
                \De_{-40, 84, -26, 3} \Big)  + 
             2\,\mb^4\,\mt^6\,
              \Big( 16 - 24\,D + 
                \sbe^4\,\De_{-72, 204, -118, 17}  
\\&{}\notag
                - 2\,\sbe^2\,\De_{-8, 48, -46, 7} \Big)  + 
             \mb^2\,\mt^8\,
              \Big( \sbe^2\,
                 \left( 13\,D^3 + \De_{-320, 292, -68} \right)  + 
                \sbe^4\,
                 \left( -22\,D^3 + \De_{304, -328, 132} \right)   
\\&{}\notag
            + \De_{16, 24, -52, 6} \Big)  + 
             \mb^8\,\mt^2\,
              \left( 3\,{(D - 2) }^2\,D + 
                \sbe^4\,\De_{-304, 328, -132, 22} + 
                \sbe^2\,\De_{288, -364, 196, -31} \right)  
\Bigg)  \Big] \Bigg\} 
\end{align}

\removelastskip

\begin{align}
\notag
&+ A_0(\MA)\,\Bigg\{ 
      - A_0(\mt) \frac{1}
{\cbe^2\,\left( \MA - 2\,\mt \right) \,
         {\left( \MA + \mb - \mt \right) }^2\,\mt^2\,
         {\left( \MA - \mb + \mt \right) }^2}
\\&{}\notag
         \frac{1}{{\left( -\MA + \mb + \mt \right) }^2\,
         {\left( \MA + \mb + \mt \right) }^2\,
         \left( \MA + 2\,\mt \right) \,
         \left( -\mb^2 + \mt^2 \right) \,\sbe^2} 
\\&{}\notag
\Bigg\{ \MA^{10}\,
            \Bigg( -{(D - 2) }^2\,D\,\mb^4\,
                \sbe^4 + 
              \mb^2\,\mt^2\,
               \left( -3\,{(D - 2) }^2\,D + 
                 6\,{(D - 2) }^2\,D\,\sbe^2 - 
                 2\,\sbe^4\,\De_{-24, 20, -2, 1} \right)   
\\&{}\notag
             + \cbe^4\,\mt^4\,\De_{16, 12, -24, 3} \Bigg)  + 
           2\,\MA^8\,\Bigg( 2\,{(D - 2) }^2\,D\,
               \mb^6\,\sbe^4 + 
              \mb^4\,\mt^2\,
               \Big( 6\,{(D - 2) }^2\,D - 
                 10\,{(D - 2) }^2\,D\,\sbe^2 
\\&{}\notag
                + \sbe^4\,\De_{-72, 68, -14, 7} \Big)  + 
              \mb^2\,\mt^4\,
               \left( \sbe^4\,
                  \left( 6\,D^3 + \De_{-192, 136, 16} \right)  - 
                 4\,\sbe^2\,
                  \left( 5\,D^3 + \De_{-4, 8, -5} \right)  + 
                 \De_{-24, 12, -2, 9} \right)  
\\&{}\notag
             - \cbe^4\,\mt^6\,\De_{88, 20, -94, 15} \Bigg)  
+ 4\,\mt^2\,{\left( \mb^2 - \mt^2 \right) }^2\,
            \Bigg( {(D - 2) }^2\,D\,\mb^8\,
               \sbe^4 + 
              \mb^6\,\mt^2\,
               \Big( 2\,{(D - 2) }^2\,
                  \left( 1 + 2\,D \right)   
\\&{}\notag
                - 4\,\sbe^2\,\De_{8, -2, -3, 1} - 
                 \sbe^4\,\De_{8, 52, -46, 5} \Big)  - 
              2\,\cbe^4\,\mt^8\,\De_{20, -4, -13, 3} + 
              \mb^4\,\mt^4\,
               \Big( -4\,\sbe^2\,\De_{-52, 112, -45, 3}  
\\&{}\notag
                + \sbe^4\,\De_{-24, 100, -54, 1} - 
                 2\,\De_{44, 20, -39, 7} \Big)  + 
              \mb^2\,\mt^6\,
               \Big( \sbe^4\,
                  \left( 9\,D^3 + \De_{72, -60, -14} \right)  + 
                 4\,\sbe^2\,\De_{-64, 114, -35, 1}  
\\&{}\notag
                + \De_{120, -360, 166, -16} \Big)  \Bigg)  - 
           2\,\MA^6\,\Bigg( 3\,{(D - 2) }^2\,D\,
               \mb^8\,\sbe^4 + 
              \mb^2\,\mt^6\,
               \Big( \sbe^4\,
                  \left( 17\,D^3 + \De_{-648, 588, 2} \right)   
\\&{}\notag
               +  \sbe^2\,
                  \left( -44\,D^3 + \De_{256, -288, 8} \right)  + 
                 \De_{-192, 36, 92, 9} \Big)  + 
              \mb^6\,\mt^2\,
               \Big( 9\,{(D - 2) }^2\,D - 
                 12\,{(D - 2) }^2\,D\,\sbe^2  
\\&{}\notag
               +  \sbe^4\,\De_{-88, 68, -10, 11} \Big)  + 
              \mb^4\,\mt^4\,
               \Big( \sbe^4\,
                  \left( 17\,D^3 + \De_{-248, 108, 78} \right)  - 
                 4\,\sbe^2\,
                  \left( 10\,D^3 + \De_{20, -16, -5} \right)  
\\&{}\notag
                + \De_{-24, 32, -46, 30} \Big)  - 
              2\,\cbe^4\,\mt^8\,\De_{148, 4, -131, 24} 
\Bigg)  + 2\,\MA^4\,\Bigg( 2\,{(D - 2) }^2\,D\,
               \mb^{10}\,\sbe^4 + 
              \mb^4\,\mt^6\,
               \Big( 4\,\sbe^2\,
                  \Big( 19\,D^3 
\\&{}\notag
              + \De_{132, 64, -143} \Big)  - 
                 2\,\sbe^4\,
                  \left( 37\,D^3 + \De_{168, 236, -298} \right)  + 
                 \De_{-328, 84, 114, 3} \Big)  + 
              \mb^8\,\mt^2\,
               \Big( 6\,{(D - 2) }^2\,D 
\\&{}\notag
               -  6\,{(D - 2) }^2\,D\,\sbe^2 + 
                 \sbe^4\,\De_{-56, 20, 6, 7} \Big)  + 
              \mb^2\,\mt^8\,
               \Big( 2\,\sbe^4\,
                  \left( 49\,D^3 + \De_{-304, 612, -248} \right)  
\\&{}\notag
                - 4\,\sbe^2\,
                  \left( 38\,D^3 + \De_{44, 152, -139} \right)  + 
                 \De_{-8, -164, 82, 25} \Big)  + 
              \mb^6\,\mt^4\,
               \Big( 4\,\sbe^4\,
                  \left( 9\,D^3 + \De_{-24, 52, -14} \right)   
\\&{}\notag
                - 4\,\sbe^2\,
                  \left( 14\,D^3 + \De_{52, 16, -41} \right)  + 
                 \De_{8, 20, -66, 35} \Big)  - 
              \cbe^4\,\mt^{10}\,\De_{440, -36, -342, 69}
              \Bigg)  
\\&{}\notag
     - \MA^2\,\Big[ {(D - 2) }^2\,D\,\mb^{12}\,
               \sbe^4 + 
              2\,\mb^6\,\mt^6\,
               \Big( 4\,\sbe^4\,
                  \left( 7\,D^3 + \De_{-104, 232, -102} \right)  + 
                 2\,\sbe^2\,
                  \left( 59\,D^3 + \De_{432, 4, -276} \right)  
\\&{}\notag
                + \De_{-368, -276, 408, -65} \Big)  + 
              \mb^{10}\,\mt^2\,
               \left( 3\,{(D - 2) }^2\,D - 
                 2\,{(D - 2) }^2\,D\,\sbe^2 + 
                 8\,\sbe^4\,\De_{-4, 2, -1, 1} \right)   
\\&{}\notag
             + \mb^8\,\mt^4\,
               \left( \sbe^4\,
                  \left( -7\,D^3 + \De_{32, -268, 244} \right)  - 
                 2\,\sbe^2\,
                  \left( 47\,D^3 + \De_{192, 156, -212} \right)  + 
                 \De_{64, 44, -124, 47} \right)  
\\&{}\notag
              + \mb^4\,\mt^8\,
               \left( \sbe^4\,
                  \left( -221\,D^3 + \De_{1088, -2900, 1380} \right)  
             + 4\,\sbe^2\,\De_{-16, -196, 140, 1} + 
                 \De_{160, 1272, -1040, 142} \right)   
\\&{}\notag
              - \cbe^4\,\mt^{12}\,\De_{608, -92, -428, 93} + 
              \mb^2\,\mt^{10}\,
               \Big( \sbe^2\,
                  \left( -330\,D^3 + \De_{-2496, 1272, 968} \right)   
\\&{}\notag
                + 8\,\sbe^4\,
                  \left( 32\,D^3 + \De_{44, 150, -153} \right)  + 
                 \De_{1120, -1892, 444, 31} \Big)  \Big]
                 \Bigg\}
 \\&{}\notag
    + A_0(\mb) \frac{1}
{\cbe^2\,\mb^2\,\left( \MA^2 - 4\,\mb^2 \right) \,
        {\left( \MA + \mb - \mt \right) }^2\,
        {\left( \MA - \mb + \mt \right) }^2\,
        {\left( -\MA + \mb + \mt \right) }^2}
\\&{}\notag
        \frac{1}{{\left( \MA + \mb + \mt \right) }^2\,
        \left( \mb^2 - \mt^2 \right) \,\sbe^2} 
\Bigg\{ \MA^{10}\,
           \Bigg( \cbe^4\,{(D - 2) }^2\,D\,
              \mt^4 - \mb^4\,\sbe^4\,
              \De_{16, 12, -24, 3} 
\\&{}\notag
            + \mb^2\,\mt^2\,
              \left( \De_{-48, 28, 8, -1} + 
                2\,\sbe^4\,\De_{-24, 20, -2, 1} + 
                2\,\sbe^2\,\De_{48, -28, -8, 1} \right)  \Bigg)  
           + 4\,{\left( \mb^3 - \mb\,\mt^2 \right) }^2
\\&{}\notag
           \Bigg( -\cbe^4\,{(D - 2) }^2\,D\,
               \mt^8 + \mb^4\,\mt^4\,
              \left( -2\,\sbe^2\,
                  \left( 5\,D^3 + \De_{-80, 124, -36} \right)  + 
                \De_{-96, 388, -204, 25} - 
                \sbe^4\,\De_{-24, 100, -54, 1} \right)   
\\&{}\notag
            + 2\,\mb^8\,\sbe^4\,\De_{20, -4, -13, 3} + 
             \mb^2\,\mt^6\,
              \left( \sbe^4\,
                 \left( 5\,D^3 + \De_{8, 52, -46} \right)  - 
                2\,\sbe^2\,
                 \left( 7\,D^3 + \De_{24, 48, -52} \right)  + 
                \De_{32, 36, -44, 5} \right)   
\\&{}\notag
            + \mb^6\,\mt^2\,
              \left( \sbe^4\,
                 \left( -9\,D^3 + \De_{-72, 60, 14} \right)  + 
                2\,\sbe^2\,
                 \left( 11\,D^3 + \De_{-56, 168, -84} \right)  + 
                \De_{64, -36, -12, 3} \right)  \Bigg)   
\\&{}\notag
         + 2\,\MA^8\,\Bigg( -
              2\,\cbe^4\,{(D - 2) }^2\,D\,\mt^6 - 
             \mb^4\,\mt^2\,
              \Big( \sbe^4\,
                 \left( 6\,D^3 + \De_{-192, 136, 16} \right)  + 
                4\,\sbe^2\,
                 \left( 2\,D^3 + \De_{92, -60, -13} \right)   
\\&{}\notag
                + \De_{-200, 116, 34, -5} \Big)  + 
             \mb^2\,\mt^4\,
              \left( \sbe^4\,
                 \left( -7\,D^3 + \De_{72, -68, 14} \right)  + 
                4\,\sbe^2\,\De_{-36, 24, 3, 1} + 
                \De_{72, -52, -2, -3} \right)   
\\&{}\notag
            + \mb^6\,\sbe^4\,\De_{88, 20, -94, 15} \Bigg)  - 
          2\,\MA^6\,\Bigg( -
              3\,\cbe^4\,{(D - 2) }^2\,D\,\mt^8 - 
             \mb^6\,\mt^2\,
              \Big( \sbe^4\,
                 \left( 17\,D^3 + \De_{-648, 588, 2} \right)   
\\&{}\notag
               + 2\,\sbe^2\,
                 \left( 5\,D^3 + \De_{520, -444, -6} \right)  + 
                \De_{-584, 336, 102, -18} \Big)  - 
             \mb^4\,\mt^4\,
              \Big( \sbe^4\,
                 \left( 17\,D^3 + \De_{-248, 108, 78} \right) 
\\&{}\notag
               + 2\,\sbe^2\,
                 \left( 3\,D^3 + \De_{288, -140, -88} \right)  + 
                \De_{-352, 204, 52, 7} \Big)  + 
             \mb^2\,\mt^6\,
              \Big( -2\,\left( 4\,D^3 + \De_{-44, 28, 1} \right)  
\\&{}\notag
               + 2\,\sbe^2\,\De_{-88, 44, 14, 5} + 
                \sbe^4\,\De_{88, -68, 10, -11} \Big)  + 
             2\,\mb^8\,\sbe^4\,\De_{148, 4, -131, 24} 
         \Bigg)  + 2\,\MA^4\,\Bigg( -
              2\,\cbe^4\,{(D - 2) }^2\,D\,\mt^{10} 
\\&{}\notag
           - \mb^2\,\mt^8\,\left( \sbe^4\,
                 \left( 7\,D^3 + \De_{-56, 20, 6} \right)  - 
                4\,\sbe^2\,
                 \left( 2\,D^3 + \De_{-28, 4, 9} \right)  + 
                \De_{-56, 20, 6, 7} \right)  
\\&{}\notag
            + \mb^6\,\mt^4\,
              \left( -4\,\sbe^2\,
                  \left( 18\,D^3 + \De_{36, 172, -155} \right)  + 
                2\,\sbe^4\,
                 \left( 37\,D^3 + \De_{168, 236, -298} \right)  + 
                \De_{136, 132, -138, -5} \right)   
\\&{}\notag
           +  \mb^4\,\mt^6\,
              \left( 4\,\sbe^2\,
                 \left( 4\,D^3 + \De_{-100, 88, 13} \right)  - 
                4\,\sbe^4\,
                 \left( 9\,D^3 + \De_{-24, 52, -14} \right)  + 
                \De_{296, -164, -42, -15} \right)   
\\&{}\notag
            + \mb^{10}\,\sbe^4\,\De_{440, -36, -342, 69} + 
             \mb^8\,\mt^2\,
              \Big( 4\,\sbe^2\,
                 \left( 11\,D^3 + \De_{-348, 460, -109} \right)  - 
                2\,\sbe^4\,
                 \left( 49\,D^3 + \De_{-304, 612, -248} \right)  
\\&{}\notag
               + \De_{792, -452, -142, 29} \Big)  \Bigg)  + 
          \MA^2\,\Bigg( \cbe^4\,{(D - 2) }^2\,D\,
              \mt^{12} + \mb^{10}\,\mt^2\,
              \Big( 8\,\sbe^4\,
                 \left( 32\,D^3 + \De_{44, 150, -153} \right)   
\\&{}\notag
               + \sbe^2\,
                 \left( -182\,D^3 + \De_{1792, -3672, 1480} \right)  + 
                \De_{-1024, 580, 188, -43} \Big)  + 
             \mb^{12}\,\sbe^4\,\De_{-608, 92, 428, -93}  
\\&{}\notag
           +  \mb^2\,\mt^{10}\,
              \left( -2\,\sbe^2\,
                  \left( 7\,D^3 + \De_{-32, 12, -4} \right)  + 
                \De_{-32, 20, -12, 9} + 
                8\,\sbe^4\,\De_{-4, 2, -1, 1} \right)  
\\&{}\notag
             + 2\,\mb^6\,\mt^6\,
              \left( 4\,\sbe^4\,
                 \left( 7\,D^3 + \De_{-104, 232, -102} \right)  - 
                2\,\sbe^2\,
                 \left( 87\,D^3 + \De_{16, 932, -684} \right)  + 
                \De_{80, 660, -552, 81} \right)  
\\&{}\notag
             + \mb^4\,\mt^8\,
              \left( \sbe^4\,\De_{32, -268, 244, -7} + 
                4\,\sbe^2\,\De_{80, 212, -228, 27}  
                - 2\,\De_{144, 268, -272, 27} \right)  
\\&{}\notag
             + \mb^8\,\mt^4\,
              \left( 2\,\sbe^2\,
                 \left( 219\,D^3 + \De_{-1056, 3292, -1660} \right)  + 
                \sbe^4\,
                 \Big( -221\,D^3 + \De_{1088, -2900, 1380} \right)   
\\&{}\notag
                + \De_{1184, -2412, 900, -75} \Big)  \Bigg)  \Bigg\}
                                                                \Bigg\}
\end{align}

\removelastskip

\begin{align}
\notag
&+ \,T_{134}(\mb^2,\mb^2,\MA^2) \, \frac{8}{\cbe^2\,
     \left( \MA^2 - 4\,\mb^2 \right) \,
     {\left( \mb^2 - \mt^2 \right) }^2\,\sbe^2}
         \Bigg\{ -
        \MA^4\,\Big( \cbe^4\,(D - 2) \, \mt^4 
\\&{}\notag
         - 4\,\left( -1 + D \right) \,\mb^4\,
            \sbe^4 + \mb^2\,\mt^2\,\sbe^2\,
            \left( 4 - 2\,D + 
              \left( -6 + 5\,D \right) \,\sbe^2 \right)  \Big)  + 
       \MA^2\,\mb^2\,\Big( (D - 2) \,\mt^4
\\&{}\notag
           \left( 6 - 8\,\sbe^2 + D\,\sbe^4 \right)  - 
          2\,\mb^2\,\mt^2\,\sbe^2\,
           \left( 6\,(D - 2)  + 
             \sbe^2\,\De_{22, -21, 2} \right)  + 
          \mb^4\,\sbe^4\,\De_{28, -32, 3} \Big)   
\\&{}\notag
       + \mb^4\,\Bigg( \mb^4\,\sbe^4\,
           \De_{-36, 48, -11, 1} - 
          2\,\mb^2\,\mt^2\,\sbe^2\,
           \left( 4 - 4\,D + \sbe^2\,\De_{-16, 30, -9, 1} 
           \right)    
\\&{}\notag
            + \mt^4\,\Big( -8\,(D - 2) 
             8\,\left( -3 + D \right) \,\sbe^2 + 
             \sbe^4\,\De_{4, 12, -7, 1} \Big)  \Bigg)  
\Bigg\} 
\\&{}\notag
  +\frac{4\,\mb^2}{\left( 4\,\mb^2 - 
       \mh^2 \right) \,\left( \mb^2 - \mt^2 \right) }
     T_{134}(\mh^2,\mb^2,\mb^2)
     \Bigg\{ \mh^2\,
        \left( \left( 2 + D \right) \,\mh^2 + 
          {(D - 4) }^2\,\mt^2 \right)  
\\&{}\notag
      + \mb^2\,\left( \mh^2\,\De_{-28, -6, 1} - 
          2\,\mt^2\,\De_{26, -5, -4, 1} \right)  + 
       2\,\mb^4\,\De_{34, 15, -8, 1} \Bigg\} 
\\&{}\notag
+ \frac{4 \, T_{134}(\mt^2,\mb^2,0)}{\mh^2}
\Bigg\{ -12\,\mb^4 + 
       \mt^2\,\left( -12\,\mt^2 + 
          \mh^2\,\De_{-22, 19, -7, 1} \right)  + 
       \mb^2\,\left( 24\,\mt^2 + 
          \mh^2\,\De_{-22, 19, -7, 1} \right)  \Bigg\} 
\\&{}\notag
  + \frac{4 \, T_{134}(\mt^2,\mb^2,\MA^2)}{\cbe^2\,
     {\left( \MA + \mb - \mt \right) }^2\,
     {\left( \MA - \mb + \mt \right) }^2\,
     {\left( -\MA + \mb + \mt \right) }^2\,
     {\left( \MA + \mb + \mt \right) }^2\,
     {\left( \mb^2 - \mt^2 \right) }^2\,\sbe^2}
\\&{}\notag
 \Bigg\{ 2\,\MA^{10}\,
        \Bigg( \cbe^4\,\left( -2 + 3\,D \right) \,\mt^4 + 
          \left( -2 + 3\,D \right) \,\mb^4\,\sbe^4 + 
          \mb^2\,\mt^2\,
           \Big( 2 - 3\,D + 2\,\left( -6 + 5\,D \right) \,
              \sbe^2 
\\&{}\notag
- 2\,\left( -6 + 5\,D \right) \,\sbe^4 
\Big)  \Bigg)  - 2\,\MA^8\,
        \Bigg( \cbe^4\,\mt^6\,\De_{-2, 7, 2} + 
          \mb^6\,\sbe^4\,\De_{-2, 7, 2} - 
          \mb^2\,\mt^4\,
           \Big( \sbe^4\,\De_{-34, 23, 2} - 
             8\,\sbe^2\,\De_{-2, -1, 1} 
\\&{}\notag
         + 4\,\De_{2, -3, 1} \Big)  + \mb^4\,\mt^2\,
           \left( \left( -52 + 54\,D - 4\,D^2 \right) \,\sbe^2 + 
             \left( 34 - 23\,D - 2\,D^2 \right) \,\sbe^4 + 
             \De_{10, -19, 2} \right)  \Bigg)  
\\&{}\notag
      + {\left( \mb^2 - \mt^2 \right) }^4\,
        \Bigg( \cbe^4\,D\,\mt^6\,\De_{6, -7, 1} + 
          D\,\mb^6\,\sbe^4\,\De_{6, -7, 1} - 
          \mb^2\,\mt^4\,
           \Big( D\,\sbe^4\,\De_{6, -7, 1} + 
             4\,\sbe^2\,\De_{-4, 22, -9, 1} 
\\&{}\notag 
            + \De_{8, -104, 50, -6} \Big)  + 
          \mb^4\,\mt^2\,
           \left( -D\,\sbe^4\,\De_{6, -7, 1} + 
             2\,\sbe^2\,\De_{-8, 50, -25, 3} + 
             \De_{8, 10, -7, 1} \right)  \Bigg)   
\\&{}\notag
       - \MA^2\,{\left( \mb^2 - \mt^2 \right) }^3\,
        \Bigg( -\cbe^4\,\mt^6\,\De_{-8, 22, -25, 3} + 
          \mb^6\,\sbe^4\,\De_{-8, 22, -25, 3} + 
          \mb^2\,\mt^4\,
           \Big( 4\,\left( -2 + 5\,D \right)   
\\&{}\notag
            - 4\,D\,\sbe^2\,\De_{56, -25, 3} + 
             \sbe^4\,\De_{-8, 154, -75, 9} \Big)  + 
          \mb^4\,\mt^2\,
           \Big( 2\,\sbe^2\,
              \left( 3\,D^3 + \De_{-8, 42, -25} \right)  + 
             \sbe^4\,\left( -9\,D^3 + \De_{8, -154, 75}
             \right)  
\\&{}\notag
+ \De_{16, 50, -25, 3} \Big)  \Bigg)  + 
       \MA^4\,\left( \mb^2 - \mt^2 \right) \,
        \Bigg( \mb^2\,\mt^6\,
           \Big( 2\,\sbe^4\,
              \left( 3\,D^3 + \De_{12, 32, -25} \right)  + 
             \sbe^2\,\left( -18\,D^3 + 
                \De_{24, -328, 158} \right)  
\\&{}\notag
            + \De_{-44, 168, -75, 9} \Big)  - 
          \cbe^4\,\mt^8\,\De_{-20, 24, -33, 3} + 
          \mb^8\,\sbe^4\,\De_{-20, 24, -33, 3} - 
          \mb^6\,\mt^2\,
           \Big( 2\,\sbe^2\,
              \left( 3\,D^3 + \De_{-36, 100, -29} \right)   
\\&{}\notag
            + 2\,\sbe^4\,
              \left( 3\,D^3 + \De_{12, 32, -25} \right)  + 
             \De_{4, -96, 33, -3} \Big)  + 
          \mb^4\,\mt^4\,\left( -1 + 2\,\sbe^2 \right) \,
           \De_{36, 112, -75, 9} \Bigg)  
\\&{}\notag
      - \MA^6\,\Bigg( \cbe^4\,
           \left( -12 - 19\,D^2 + D^3 \right) \,\mt^8 + 
          \left( -12 - 19\,D^2 + D^3 \right) \,\mb^8\,\sbe^4 - 
          \mb^4\,\mt^4\,
           \Big( \De_{4, 32, -19, 1}  
\\&{}\notag
            - 2\,\sbe^2\,\De_{36, 64, -47, 5} + 
             2\,\sbe^4\,\De_{36, 64, -47, 5} \Big)  + 
          \mb^2\,\mt^6\,
           \left( 4\,\sbe^4\,\De_{-20, 32, -7, 1} - 
             2\,\sbe^2\,\De_{-4, -8, 5, 1} + 
             \De_{44, -56, 19, -1} \right)   
\\&{}\notag
         + \mb^6\,\mt^2\,
           \left( \De_{-28, 88, -19, 1} + 
             4\,\sbe^4\,\De_{-20, 32, -7, 1} + 
             \sbe^2\,\De_{152, -272, 66, -6} \right)  \Bigg)  
\Bigg\}
\\&{}\notag
  + \frac{4\, \,
     T_{134}(\mt^2,\mh^2,\mb^2)}{\mh^2} \Bigg\{ 12\,\mb^4 + 
       \left( \left( 2 + D \right) \,\mh^2 - 12\,\mt^2 \right) \,
        \left( \mh^2 - \mt^2 \right)  - 
       \mb^2\,\left( \left( 14 + D \right) \,\mh^2 + 
          24\,\mt^2 \right)  \Bigg\}
\\&{}\notag
 + \frac{8}{\cbe^2\,
     \left( \MA - 2\,\mt \right) \,\left( \MA + 2\,\mt \right) \,
     {\left( \mb^2 - \mt^2 \right) }^2\,\sbe^2}
     T_{134}(\mt^2,\mt^2,\MA^2)
 \Bigg\{ \MA^4\,\Big( 4\,\cbe^4\,
           \left( -1 + D \right) \,\mt^4 
\\&{}\notag
         - (D - 2) \,\mb^4\,\sbe^4 + 
          \cbe^2\,\mb^2\,\mt^2\,
           \left( 2 - 3\,D + \left( -6 + 5\,D \right) \,\sbe^2 
                  \right)  \Big)  
\\&{}\notag
     + \MA^2\,\mt^2\,
        \Bigg( 2\,\cbe^2\,\mb^2\,\mt^2\,
           \left( \left( 22 - 21\,D + 2\,D^2 \right) \,\sbe^2 + 
             \De_{-10, 15, -2} \right)  + 
          \mb^4\,\Big( {(D - 2) }^2 + 
             (D - 2) \,D\,\sbe^4 
\\&{}\notag
             - 2\,\sbe^2\,\De_{8, -6, 1} \Big)  + 
          \cbe^4\,\mt^4\,\De_{28, -32, 3} \Bigg)  + 
       \mt^4\,\Bigg( \cbe^4\,\mt^4\,
           \De_{-36, 48, -11, 1} + 
          \mb^4\,\Big( -2\,\sbe^2\,\De_{-8, 16, -7, 1} 
\\&{}\notag
            + \De_{-4, 12, -7, 1} + \sbe^4\,\De_{4, 12, -7, 1}
 \Big)  + 2\,\cbe^2\,\mb^2\,\mt^2\,
           \left( \sbe^2\,\De_{-16, 30, -9, 1} + 
             \De_{12, -26, 9, -1} \right)  \Bigg)  \Bigg\} 
\\&{}\notag
+  \frac{4\,\mt^2}{\left( \mh - 2\,\mt 
\right) \,\left( \mh + 2\,\mt \right) \,
     \left( -\mb^2 + \mt^2 \right) }
     T_{134}(\mt^2,\mt^2,\mh^2)
\Bigg\{ \left( 2 + D \right) \,\mh^4 + 
       \mh^2\,\mt^2\,\De_{-28, -6, 1}  
\\&{}\notag
      + \mb^2\,\left( {(D - 4) }^2\,\mh^2 - 
          2\,\mt^2\,\De_{26, -5, -4, 1} \right)  + 
       2\,\mt^4\,\De_{34, 15, -8, 1} \Bigg\} \; \Bigg]
\end{align}

\end{appendix}



\begin{thebibliography}{00}  

\bibitem{susy} H.P.~Nilles, 
               {\em Phys.\ Rep.} {\bf 110} (1984) 1; \\ 
               H.E.~Haber and G.L.~Kane, 
               {\em Phys.\ Rep.} {\bf 117}, (1985) 75; \\  
               R.~Barbieri, 
               {\em Riv.\ Nuovo Cim.} {\bf 11}, (1988) 1. 

\bibitem{pdg} Part. Data Group,
              {\em Phys. Rev.} {\bf D 66} (2002) 010001.

\bibitem{hhg} J.~Gunion, H.~Haber, G.~Kane and S.~Dawson,
              {\em The Higgs Hunter's Guide}, Addison-Wesley, 1990.

\bibitem{lephiggs} The LEP working group for Higgs boson searches, 
                   LHWG Note 2001-4; LHWG Note 2001-5, 
                   see {\tt lephiggs.web.cern.ch/LEPHIGGS/papers/}.

\bibitem{rho} M.~Veltman, 
              {\em Nucl. Phys.} {\bf B 123} (1977) 89. 

\bibitem{dr1lA} R.~Barbieri and L. Maiani,  
                {\em Nucl. Phys.} {\bf B 224} (1983) 32; \\
                C.~S.~Lim, T.~Inami and N.~Sakai, 
                {\em Phys. Rev.} {\bf D 29} (1984) 1488; \\
                E.~Eliasson, 
                {\em Phys. Lett.} {\bf B 147} (1984) 65; \\
                Z.~Hioki, 
                {\em Prog. Theo. Phys.} {\bf 73} (1985) 1283; \\
                J.~A.~Grifols and J.~Sola, 
                {\em Nucl. Phys.} {\bf B 253} (1985) 47; \\
                B.~Lynn, M.~Peskin and R.~Stuart, 
                CERN Report 86-02, p. 90; \\
                R.~Barbieri, M.~Frigeni, F.~Giuliani and H.E.~Haber, 
                {\em Nucl. Phys.} {\bf B 341} (1990) 309; \\   
                M.~Drees and K.~Hagiwara, 
                {\em Phys. Rev.} {\bf D 42} (1990) 1709.

\bibitem{dr1lB} M.~Drees, K.~Hagiwara and A.~Yamada, 
                {\em Phys. Rev.} {\bf D 45} (1992) 1725; \\ 
                P.~Chankowski, A.~Dabelstein, W.~Hollik, W.~M\"osle, 
                S.~Pokorski and J.~Rosiek, 
                {\em Nucl. Phys.} {\bf B 417} (1994) 101;\\ 
                D.~Garcia and J.~Sol\`a, 
                {\em Mod. Phys. Lett.} {\bf A 9} (1994) 211.


\bibitem{dr2lA} A.~Djouadi, P.~Gambino, S.~Heinemeyer, W.~Hollik,
                C.~J\"unger and G.~Weiglein, 
                {\em Phys. Rev. Lett.} {\bf 78} (1997) 3626,
                hep-ph/9612363;
                {\em Phys. Rev.} {\bf D 57} (1998) 4179,
                hep-ph/9710438.

\bibitem{dr2lB} S.~Heinemeyer, 
                PhD thesis, 
                see {\tt www-itp.physik.uni-karlsruhe.de/prep/phd/};\\
                G.~Weiglein, 
                hep-ph/9901317;\\
                S.~Heinemeyer, W.~Hollik and G.~Weiglein,
                {\em in preparation}.
                
\bibitem{decoupling1l} T.~Appelquist and J.~Carazzone,
                       {\em Phys. Rev.} {\bf D 11} (1975) 2856;\\
                       A.~Dobado, M.~Herrero and S.~Pe\~naranda,
                       {\em Eur. Phys. Jour.} {\bf C 7} (1999) 313,
                       hep-ph/9710313,
                       {\em Eur. Phys. Jour.} {\bf C 12} (2000) 673,
                       hep-ph/9903211;
                       {\em Eur. Phys. Jour.} {\bf C~17} (2000) 487,
                       hep-ph/0002134.

\bibitem{drMSSMgf2} S.~Heinemeyer and G.~Weiglein, 
                    proceedings of the RADCOR2000, Carmel, Sep.~2000,
                    hep-ph/0102317.

\bibitem{drSMgf2mt4} R.~Barbieri, M.~Beccaria, P.~Ciafaloni, G.~Curci 
                     and A.~Vicere,
                     {\em Nucl. Phys.} {\bf B 409} (1993) 105;\\
                     J.~Fleischer, F.~Jegerlehner and O.V.~Tarasov,
                     {\em Phys. Lett.} {\bf B 319} (1993) 249.

\bibitem{drSMgfals} A.~Djouadi and C.~Verzegnassi,
                    {\em Phys. Lett.} {\bf B 195} (1987) 265;\\
                    A.~Djouadi,
                    {\em Nuovo Cim.} {\bf A 100} (1988) 357.

\bibitem{drSMgfals2} K.~Chetyrkin, J.H.~K\"uhn and M.~Steinhauser,
                     {\em Phys. Rev. Lett.} {\bf 75} (1995) 3394,
                     hep-ph/9504413;\\
                     L.~Avdeev et al.,
                     {\em Phys. Lett.} {\bf B 336} (1994) 560,
                     hep-ph/9406363;
                     E: {\em Phys. Lett.} {\bf B 349} (1995) 597.

\bibitem{drSMgf2mh0} J.~Van~der~Bij and F.~Hoogeveen,
                     {\em Nucl. Phys.} {\bf B 283} (1987) 477.

\bibitem{drSMgf3mh0} J.~Van der Bij, K.~Chetyrkin, M.~Faisst, G.~Jikia 
                     and T.~Seidensticker,
                     {\em Phys. Lett.} {\bf B~498} (2001) 156,
                     hep-ph/0011373. 

\bibitem{mhiggslong} S.~Heinemeyer, W.~Hollik and G.~Weiglein, 
                     {\em Eur. Phys. Jour.} {\bf C 9} (1999) 343, 
                     hep-ph/9812472.

\bibitem{feynarts} J.~K\"ublbeck, M.~B\"ohm and A.~Denner, 
                   {\em Comp. Phys. Comm.} {\bf 60} (1990) 165;\\
                   T.~Hahn and M.~Perez-Victoria,
                   {\em Comput. Phys. Commun.} {\bf 118} (1999) 153, 
                   hep-ph/9807565;\\
                   T.~Hahn,
                   {\em Nucl. Phys. Proc. Suppl.} {\bf 89} (2000) 231, 
                   hep-ph/0005029;
                   {\em Comput. Phys. Commun.} {\bf 140} (2001) 418, 
                   hep-ph/0012260.\\
                   The program is available via {\tt www.feynarts.de} .

\bibitem{famssm} T.~Hahn and C.~Schappacher,
                 {\em Comput. Phys. Commun.} {\bf 143} (2002) 54,
                 hep-ph/0105349.

\bibitem{2lred} G.~Weiglein, R.~Scharf and M.~B\"ohm,
                {\em Nucl. Phys.} {\bf B 416} (1994) 606,
                hep-ph/9310358;\\
                G.~Weiglein, R.~Mertig, R.~Scharf and M.~B\"ohm, 
                in {\it New Computing Techniques in Physics Research 2},
                ed.~D.~Perret-Gallix (World Scientific, Singapore,
                1992), p.~617.

\bibitem{a0b0c0d0} G.~Passarino and  M.~Veltman, 
                   {\em Nucl. Phys.} {\bf B 160} (1979) 151.

\bibitem{t134} A. Davydychev und J. B. Tausk, 
               {\em Nucl. Phys.} {\bf B 397} (1993) 123;\\
               F. Berends und J. B. Tausk,
               {\em Nucl. Phys.} {\bf B 421} (1994) 456.

\bibitem{mhiggs1l} H.~Haber and R.~Hempfling,
                   {\em Phys. Rev. Lett.} {\bf 66} (1991) 1815;\\
                   Y.~Okada, M.~Yamaguchi and T.~Yanagida,
                   {\em Prog. Theor. Phys.} {\bf 85} (1991) 1;\\
                   J.~Ellis, G.~Ridolfi and F.~Zwirner,
                   {\em Phys. Lett.} {\bf B 257} (1991) 83; 
                   {\em Phys. Lett.} {\bf B 262} (1991) 477;\\
                   R.~Barbieri and M.~Frigeni,
                   {\em Phys. Lett.} {\bf B 258} (1991) 395.

\bibitem{mhiggsAEC} G.~Degrassi, S.~Heinemeyer, W.~Hollik,
                    P.~Slavich and G.~Weiglein, 
                    {\em in preparation}.

\bibitem{feynhiggs} S.~Heinemeyer, W.~Hollik and G.~Weiglein, {\em
                    Comp. Phys. Comm.} {\bf 124} 2000 76,
                    hep-ph/9812320; \\
                    M.~Frank, S.~Heinemeyer, W.~Hollik and G.~Weiglein,
                    hep-ph/0202166.\\
                    The code is accessible via
                    {\tt www.feynhiggs.de} .

\bibitem{LHbenchmark} M.~Carena, S.~Heinemeyer, C.~Wagner and G.~Weiglein,
                      hep-ph/0202167.

\bibitem{tbexcl} S.~Heinemeyer, W.~Hollik and G.~Weiglein, 
                 {\em JHEP} {\bf 0006} (2000) 009, 
                 hep-ph/9909540;\\
                 A.~Dedes, S.~Heinemeyer, P.~Teixeira-Dias 
                 and G.~Weiglein, 
                 {\em Jour. Phys.}  {\bf G 26} (2000) 582,
                 hep-ph/9912249.

\bibitem{mhiggsletter} S.~Heinemeyer, W.~Hollik and G.~Weiglein, 
                       {\em Phys. Rev.} {\bf D 58} (1998) 091701, 
                       hep-ph/9803277; 
                       {\em Phys. Lett.} {\bf B 440} (1998) 296, 
                       hep-ph/9807423;
                       hep-ph/9806250.

\bibitem{mhiggslle} S.~Heinemeyer, W.~Hollik and G.~Weiglein, 
                   {\em Phys. Lett.} {\bf B 455} (1999) 179,
                   hep-ph/9903404.

\bibitem{grueni} M.~Gr\"unewald,
                 talk given at ICHEP02, Amsterdam, July 2002, see\\
                 {\tt www.ichep02.nl/MainPages/PlenaryProgram.html} .

\bibitem{deltarferm} G.~Degrassi, P.~Gambino and A.~Vicini,
                     {\em Phys. Lett.} {\bf B 383} (1996) 219, 
                     hep-ph/9603374;\\
                     G.~Degrassi, P.~Gambino and A.~Sirlin,
                     {\em Phys. Lett.} {\bf B 394} (1997) 188
                     hep-ph/9611363;\\
                     A.~Freitas, W.~Hollik, W.~Walter and G.~Weiglein,
                     {\em Phys. Lett.} {\bf B 495} (2000) 338, 
                     hep-ph/0007091;\\
                     A.~Freitas, S.~Heinemeyer, W.~Hollik, W.~Walter 
                     and G.~Weiglein,
                     {\em Nucl. Phys. Proc. Suppl.} {\bf 89} (2000) 82,
                     hep-ph/0007129;\\
                     A.~Freitas, W.~Hollik, W.~Walter and G.~Weiglein,
                     {\em Nucl. Phys.} {\bf B 632} (2002) 189, 
                     hep-ph/0202131.

\bibitem{EWPOSM} U.~Baur, R.~Clare, J.~Erler, S.~Heinemeyer, 
                 D.~Wackeroth, G.~Weiglein and D.~R.~Wood,
                 in {\it Proc. of the APS/DPF/DPB Summer Study on the 
                         Future of Particle Physics (Snowmass 2001) } 
                 ed.\ R.~Davidson and C.~Quigg,
                 hep-ph/0111314.

\bibitem{moenig} R.~Hawkings and K.~M\"onig, 
                 {\em EPJdirect} {\bf C 8} (1999) 1,
                 hep-ex/9910022.

\bibitem{gigaz} S.~Heinemeyer, T.~Mannel and G.~Weiglein,
                hep-ph/9909538;\\ 
                J.~Erler, S.~Heinemeyer, W.~Hollik, G.~Weiglein 
                and P.M.~Zerwas,
                {\em Phys. Lett.} {\bf B 486} (2000) 125,
                hep-ph/0005024;\\
                J.~Erler and S.~Heinemeyer,
                hep-ph/0102083.

\bibitem{MarHowRep} M.~Carena and H.~Haber,
                    hep-ph/0208209.

\bibitem{EWPOMSSM} S.~Heinemeyer and G.~Weiglein,
                   {\em Nucl. Phys. Proc. Suppl.} {\bf 89} (2000) 216,
                   hep-ph/0007307.

\bibitem{LEPHiggsSM} [LEP Higgs working group], LHWG Note/2002-01,\\
                     {\tt http://lephiggs.web.cern.ch/LEPHIGGS/papers/}.

\bibitem{LHCprec} M.~Beneke et al.,
                  {\em Top Quark Physics}, in CERN 2000-004,
                  eds.\ G.~Altarelli and M.~Mangano, 
                  hep-ph/0003033; \\
                  S.~Haywood et al., 
                  {\em Electroweak physics}, in CERN 2000-004,
                  eds.\ G.~Altarelli and M.~Mangano, 
                  hep-ph/0003275.

\bibitem{topthresholdyt} M.~Martinez and R.~Miquel,
                         hep-ph/0207315.

\end{thebibliography}
\end{document}